\documentclass[aps,twocolumn,showpacs,preprintnumbers,floats]{revtex4}\def\@cite#1#2{\textsuperscript{[{#1\if@tempswa , #2\fi}]}}
\usepackage{mathrsfs}

\usepackage{longtable,lscape}
\usepackage{savesym}
\usepackage{txfonts}
\savesymbol{iint}
\savesymbol{iiint}
\savesymbol{iiiint}
\savesymbol{idotsint}
\usepackage{amssymb}
\usepackage{amsmath}
\restoresymbol{TXF}{iint}
\restoresymbol{TXF}{iiint}
\restoresymbol{TXF}{iiiint}
\restoresymbol{TXF}{idotsint}
\usepackage{indentfirst}
\usepackage{graphicx,booktabs}
\usepackage{braket}
\usepackage{multirow}
\usepackage{color}
\usepackage{subfigure}
\usepackage{float}
\usepackage{bm}
\usepackage{epsfig}
\usepackage[colorlinks, citecolor=blue,anchorcolor=red,menucolor=red, linkcolor=red,filecolor=red,urlcolor=blue,frenchlinks=red]{hyperref}
\newcommand{\vsig}{\mbox{\boldmath$\sigma$\unboldmath}}

\begin{document}
	
	\title{Probing the inner structures of the observed $\Xi_b$ and $\Xi_b'$ resonances}
	\author{Yu-Bin Zhang$^{1}$, Yi-Heng Wang$^{1}$, Hui-Hua Zhong$^{2}$~\footnote {E-mail: zhonghuihua@foxmail.com}, Li-Ye Xiao$^{1}$~\footnote {E-mail: lyxiao@ustb.edu.cn}}
\affiliation{ 1)Institute of Theoretic
al Physics, University of Science and Technology Beijing,
Beijing 100083, China}
\affiliation{ 2) College of Aeronautics Mechanical and Electrical Engineering, Jiangxi Flight University, Nanchang 330088, China }

\begin{abstract}
To shed light on the inner structure of the observed single-bottom strange baryons, in this work we systematically study the Okubo-Zweig-Iizuka allowed strong decay properties of $1P$- and $2S$-wave $\Xi_b$ and $\Xi_b'$ baryons within the $j-j$ coupling scheme in the framework of the quark pair creation model. For a comparison, we also give the predictions of the chiral quark model. The calculations indicate that: (i) The $1P$-wave $\lambda$-mode $\Xi_b$ states $\Xi_b|J^P=1/2^-,1\rangle_{\lambda}$ and $\Xi_b|J^P=3/2^-,1\rangle_{\lambda}$ are highly promising candidates for the observed state $\Xi_b(6087)$ and $\Xi_b(6095)/\Xi_b(6100)$, respectively. The $1P$-wave $\lambda$-mode $\Xi_b'$ states $\Xi_b'|J^P=3/2^-,2\rangle_{\lambda}$ and $\Xi_b'|J^P=5/2^-,2\rangle_{\lambda}$ are likely candidates for the state $\Xi_b(6227)$. Meanwhile, we cannot rule out the possibility that $\Xi_b(6227)$ could be a candidate of the $1P$-wave $\rho$-mode $\Xi_b$ state $\Xi_b|J^P=3/2^-,2\rangle_{\rho}$ or $\Xi_b|J^P=5/2^-,2\rangle_{\rho}$. (ii) For the other $1P$-wave $\rho$-mode $\Xi_b$ states and $1P$-wave $\lambda$-mode $\Xi_b'$ states, they may be moderate states with a width of
several tens of MeV. Their main decay channels are $\Xi_b\pi$, $\Xi_b'\pi$, $\Xi_b^*\pi$ or $\Lambda_b\bar{K}$. The width of the $1P$-wave $\rho$-mode $\Xi_b'$ states are slightly broader, approximately several tens to over one hundred MeV, and the dominant decay channels are $\Xi_b'\pi$, $\Xi_b^*\pi$, $\Sigma_bK$ or $\Sigma_b^*K$. (iii) The $2S$-wave $\lambda$-mode $\Xi_b$ and $\Xi_b'$ states are most likely to be relatively narrow state with a width of only a few to around ten MeV, and they mainly decay into $\Xi_b'\pi$ or $\Xi_b^*\pi$. Meanwhile, the decay  process of $2S$-wave $\lambda$-mode $\Xi_b'$ states into the final states containing $1P$-wave $\Xi_b$ baryons should not be overlooked either. We hope that the theoretical results can provide some reference for future experimental investigations.
\end{abstract}
	\maketitle

\section{Introduction}

Exploring internal structures and finding the missing resonance states is one of the important topics in hadronic physics, which will help us better understand strong interaction and the theoretical framework of  nonperturbative quantum chromodynamics(QCD). Studies of hadrons containing a heavy quark play a special role since the heavy quark symmetry can be exploited.

There are a number of $b$ baryon states that contain both beauty and strange quarks. The states form isospin doublets $\Xi_b^{(0,-)}$. In the quark model~\cite{Gell-Mann:1964ewy,Klempt:2009pi,Zweig:1964ruk}, three isospin doublets of ground states are expected. So far, five of these states have been observed in the experiment~\cite{CDF:2007cgg,CDF:2011ipk,CMS:2012frl,LHCb:2014nae}, only the $\Xi_b'^0$ baryon remains unobserved. This is presumably because its mass lies below the threshold of $\Xi_b^-\pi^+$~\cite{LHCb:2016mrc} and makes it experimentally challenging to observed. Moreover, a great progress has been achieved on the observation of excited $\Xi_b$ states. In 2018, the LHCb Collaboration firstly reported a resonance, $\Xi_b(6227)^-$, decaying into both $\Lambda_b^0K^-$ and $\Xi_b^0\pi^-$~\cite{LHCb:2018vuc}. Furthermore, the ratio of branching fractions were measured to be $B(\Xi_b(6227)^-\rightarrow\Lambda_b^0K^-)/B(\Xi_b(6227)^-\rightarrow\Xi_b^0\pi^-)\simeq1.0\pm0.5$~\cite{LHCb:2018vuc}, which provided a good reference for theoretical explanation. In 2021,  the LHCb Collaboration confirmed this state again, and at the same time reported a new beauty-baryon resonance $\Xi_b(6227)^0$ decaying into $\Xi_b^-\pi^+$~\cite{LHCb:2020xpu}. The measured masses of the $\Xi_b(6227)^-$ and $\Xi_b(6227)^0$ states are consistent with them being isospin partners. Soon after, the CMS Collaboration clearly observed another narrow new resonance $\Xi_b(6100)^-$ near the $\Xi_b^-\pi^+\pi^-$ kinematic threshold~\cite{CMS:2021rvl}. And this new resonance was later confirmed by the LHCb Collaboration in the $\Xi_b^{*0}(\Xi_b^-\pi^+)\pi^-$ mass distribution~\cite{LHCb:2023zpu}. The $\Xi_b(6100)^-$ resonance and its decay sequence were consistent with the lightest orbitally excited $\Xi_b^-$ baryon, with spin and parity quantum numbers $J^P=3/2^-$~\cite{CMS:2021rvl}. In 2022, two narrow resonant states, $\Xi_b(6327)^0$ and $\Xi_b(6333)^0$, were observed by the LHCb Collaboration again in the $\Lambda_b^0K^-\pi^+$ mass spectrum, which properties were in good agreement with the expectations for a doublet of $1D$ $\Xi_b^0$ states~\cite{LHCb:2021ssn}. Recently, the LHCb Collaboration announced again two new baryonic structures, $\Xi_b(6087)^0$ and $\Xi_b(6095)^0$, in the final state $\Xi_b^0\pi^+\pi^-$~\cite{LHCb:2023zpu}. And data indicated that the $\Xi_b(6087)^0$ baryon decayed mainly through the $\Xi_b'^-\pi^+$ state and the $\Xi_b(6095)^0$ baryon mainly through the $\Xi_b^{*-}\pi^+$ state~\cite{LHCb:2023zpu}. Up to now there are twelve $\Xi_b$ baryons observed experimentally according to the Particle Data Group~\cite{ParticleDataGroup:2024cfk} and we collect those states in Table~\ref{Data}.

For constructing the highly excited bottom baryon spectroscopy, there exist many theoretical studies in the literature. Before the experiment, various theoretical models and calculations predicted the masses and decay properties of excited $\Xi^{(')}_b$ baryons~\cite{Wang:2017kfr,Kawakami:2019hpp,Kim:2021ywp,Chen:2014nyo,Ebert:2011kk,Roberts:2007ni,Vijande:2012mk,Chen:2019ywy,Yao:2018jmc,Chen:2016phw,
Wei:2016jyk,Garcia-Recio:2012lts,Valcarce:2008dr,Chen:2019ywy,Ebert:2007nw,Jenkins:2007dm,Karliner:2008sv,Garcilazo:2007eh,Mao:2015gya,Wang:2010it,Wang:2017goq,Thakkar:2016dna,Chen:2018orb}, which established a valuable theoretical framework for experimental observation~\cite{CMS:2021rvl,LHCb:2021ssn}. After the experiment, a great deal of theoretical work was also stimulated to decode the inner structures of those newly observed states. For the heaviest observed excited states $\Xi_b(6327)^0$ and $\Xi_b(6333)^0$, the most popular explanation is interpreting them into $1D$-wave $\Xi_b$ resonances with spin-parity $J^P=3/2^+$ and $5/2^+$~\cite{Kakadiya:2022zvy,Bijker:2020tns,Garcia-Tecocoatzi:2023btk,Wang:2022zcy,Zhou:2023xxs,Li:2022xtj}, respectively. Meanwhile, there are other interpretations, such as $1P$- wave $\Xi'_b$ states with spin-parity $J^P=1/2^-$~\cite{Weng:2024roa,Bijker:2023zhh}, $3/2^-$~\cite{Weng:2024roa,Bijker:2023zhh,Jakhad:2024fin,Aliev:2018lcs} or $5/2^-$~\cite{Jakhad:2024fin}. For the first observed excited state $\Xi_b(6227)$, the theory explains it as $1P$-wave $\Xi'_b$ states~\cite{Li:2023gbo,Li:2022xtj} with spin-parity $J^P=1/2^-$~\cite{Li:2024zze}, $J^P=3/2^-$~\cite{Chen:2018orb,He:2021xrh,Weng:2024roa,Wang:2018fjm,Jakhad:2024fin,Cui:2019dzj}, or $J^P=5/2^-$~\cite{Chen:2018orb,Wang:2018fjm,Bijker:2020tns,Bijker:2023zhh,Ortiz-Pacheco:2023kjn,Garcia-Tecocoatzi:2023btk}. Interpreted $\Xi_b(6227)$ as $2S$-wave $\Xi_b$ state~\cite{Li:2022xtj} with spin-parity $J^P=1/2^+$~\cite{Kakadiya:2022zvy,Weng:2024roa} is also allowed in theory. Besides the above conventional explanations, there exist some exotic explanations for the $\Xi_b(6227)$, such as $\Sigma_b\bar{K}$ molecular~\cite{Huang:2018bed,Zhu:2020lza,Nieves:2019jhp} and dynamically generated state~\cite{Yu:2018yxl}. For the $\Xi_b(6100)^-$ and most recently observed $\Xi_b(6095)^0$ states, we can tentatively assign them to be the isospin doublet by considering the quarks constituents and mass difference. Currently, the vast majority of theoretical work interprets the two states as $1P$-wave $\Xi_b$ resonance with spin-parity $J^P=3/2^-$~\cite{He:2021xrh,Arifi:2021orx,Wang:2024rai,Ortiz-Pacheco:2023kjn,Weng:2024roa,Kakadiya:2022zvy,Tan:2023opd,Li:2023gbo,Bijker:2023zhh,Garcia-Tecocoatzi:2023btk,Xin:2023usd,Li:2022xtj}, but a very small number of studies suggest them may be $1P$-wave $\Xi_b$ resonance with spin-parity $J^P=1/2^-$~\cite{Jakhad:2024fin}. For the other most recently reported state $\Xi_b(6087)^0$, the theoretical work interprets this state as $1P$-wave $\Xi_b$ baryon with spin-parity $J^P=1/2^-$~\cite{Wang:2024rai,Ortiz-Pacheco:2023kjn,Tan:2023opd,Li:2023gbo,Bijker:2023zhh,Xin:2023usd}, which is inconsistent with the suggested spin-parity $J^P=3/2^-$ from Particle Data Group~\cite{ParticleDataGroup:2024cfk}. The more theoretical studies with more references can be found in review articles~\cite{HFLAV:2022esi,Chen:2022asf}.

According to the predicted spectra of the $\Xi_b$ and $\Xi'_b$ baryons~\cite{Ebert:2011kk,Cui:2019dzj,Yang:2020zrh,Ebert:2007nw,Kakadiya:2022zvy,Yu:2021zvl,Jia:2019bkr,Li:2022xtj,Yang:2022oog}, the mass distribution of the observed excited-state candidates lies between the predicted masses of the $1P$-, $1D$-, and $2S$-wave states. In our previous works~\cite{Wang:2022zcy,Zhou:2023xxs}, we have systematically studied the decay properties of the $1D$-wave $\Xi^{(')}_b$ baryons and given our theoretical explanations for the $\Xi_b(6327)^0$ and $\Xi_b(6333)^0$ states. In the present work, we carry out a systematical study of the $1P$- and $2S$-wave $\Xi^{(')}_b$ states for both $\lambda$- and $\rho$-mode excitations with the quark pair creation(QPC) model. Meanwhile, we also include the chiral quark model's predictions for comparison. Through studies of their strong decay properties, we attempt to clarify the internal structure of the controversial excited states: $\Xi_b(6087)^0$, $\Xi_b(6095)^0/\Xi_b(6100)^-$, and $\Xi_b(6227)^{(0,-)}$.  The quark model classification and Okubo-Zweig-Iizuka(OZI) allowed two-body decay modes are collected in Table~\ref{table2}.

\begin{table}[]
	\caption{\label{Data} Masses and decay widths of the $\Xi_b$ baryons from the Particle Data Group~\cite{ParticleDataGroup:2024cfk}. The unit is MeV.}
	\centering
	\begin{tabular}{cccc}
		\hline\hline
State     &$J^P$  &Mass \& Width &Status \\  \hline
$\Xi_b^-$ &$1/2^+$ &M=$5797.0\pm0.4$  &$\ast\ast\ast$\\
          &        &-                  & \\
$\Xi_b^0$ &$1/2^+$ &M=$5791.7\pm0.4$  &$\ast\ast\ast$\\
          &        &-                  & \\		
$\Xi_b'(5935)^-(\Xi_b'^-)$ &$1/2^+$ &M=$5934.9\pm0.4$  &$\ast\ast\ast$\\
                 &        &~~$\Gamma=0.03\pm0.032$   & \\
$\Xi_b(5945)^0(\Xi_b^{*0})$ &$3/2^+$ &M=$5952.3\pm0.6$  &$\ast\ast\ast$\\
                 &        &$\Gamma=0.87\pm0.07$   & \\
$\Xi_b(5955)^-(\Xi_b^{*-})$ &$3/2^+$ &M=$5955.5\pm0.4$  &$\ast\ast\ast$\\
                 &        &$\Gamma=1.43\pm0.11$   & \\
$\Xi_b(6087)^0$ &$3/2^-$ &M=$6087.0\pm0.5$  &$\ast\ast\ast$\\
                 &        &$\Gamma=2.4\pm0.5$~~~~~   & \\
$\Xi_b(6095)^0$ &$3/2^-$ &M=$6095.1\pm0.4$  &$\ast\ast\ast$\\
                 &        &$\Gamma=0.50\pm0.35$   & \\
$\Xi_b(6100)^-$ &$3/2^-$ &M=$6099.8\pm0.4$  &$\ast\ast\ast$\\
                 &        &$\Gamma=0.94\pm0.31$   & \\
$\Xi_b(6227)^-$ &$?^?$ &M=$6227.9\pm0.9$  &$\ast\ast\ast$\\
                &        &$\Gamma=19.9\pm2.6$~~   & \\
$\Xi_b(6227)^0$ &$?^?$ &M=$6226.8\pm1.6$  &$\ast\ast\ast$\\
                &        &$\Gamma=19^{+5}_{-4}$~~~~~~~~~~~~   & \\
$\Xi_b(6327)^0$ &$?^?$ &~~~~M=$6327.28\pm0.35$  &$\ast\ast\ast$\\
                &        &$\Gamma<2.56$~~~~~~~~~~~~   & \\
$\Xi_b(6333)^0$ &$?^?$ &~~~~M=$6332.69\pm0.28$  &$\ast\ast\ast$\\
                &        &$\Gamma<1.92$~~~~~~~~~~~~   & \\
\hline\hline
\end{tabular}
\end{table}

\begin{table*}
	\caption{Predicted mass of $\Xi_b^{'}$ and $\Xi_b$ baryons($1P$- and $2S$-wave) in various quark models and possible OZI-allowed two body decay modes.}	
	\label{table2}
	\resizebox{\textwidth}{!}{%
\begin{tabular}{cccccccccccccccccccc}
		\toprule[1.2pt]
\multirow{1}{*}{Notation}& \multicolumn{8}{c}{$\rm{Quantum}$ $\rm{Number}$}& \multicolumn{6}{c}{$\rm{Predicted}$ $\rm{Mass}$}  & \multicolumn{1}{c}{$\rm{Decay}$ $\rm{mode}$} &$\rm{Candidate}$  \\
				\hline
		$\Xi_b$${\ket{J^P,j}}_{\lambda(\rho)}$&$n_{\lambda}$&$n_{\rho}$&$l_{\lambda}$&$l_{\rho}$&$L$&$s_{\rho}$&$j$&$J^P$
&GIM\cite{Li:2022xtj}&hCQM\cite{Kakadiya:2022zvy}& NQM\cite{Ortiz-Pacheco:2023kjn}& RQM\cite{Ebert:2011kk}&HM\cite{Garcia-Tecocoatzi:2023btk}&RFTM\cite{Jakhad:2024fin}& \\		
		\hline
    $\Xi_b$${\ket{J^P=\frac{1}{2}^-,1}}_\lambda$	&0&0&1&0&1&0&1&$\frac{1}{2}^{-}$&6084&&6077&6120&6079&6112 &$\Xi'^{(*)}_b\pi$ &$\Xi_b(6087)$ \\
	$\Xi_b$${\ket{J^P=\frac{3}{2}^-,1}}_\lambda$	&0&0&1&0&1&0&1&$\frac{3}{2}^{-}$&6097&&6084&6130&6085&6126 &$\Xi'^{(*)}_b\pi$ &$\Xi_b(6095/6100)$\\
		\hline
	$\Xi_b$${\ket{J^P=\frac{1}{2}^-,0}}_{\rho}$&0&0&0&1&1&1&0&$\frac{1}{2}^{-}$&&6137&6243&&6248&  &$\Xi_b\pi$,$\Xi'^{(*)}_b\pi$,$\Xi_b(1P_{\lambda})\pi$,$\Lambda_b\bar{K}$ \\
	$\Xi_b$${\ket{J^P=\frac{1}{2}^-,1}}_{\rho}$&0&0&0&1&1&1&1&$\frac{1}{2}^{-}$ &&6138&6258&&6271&&$\Xi_b\pi$,$\Xi'^{(*)}_b\pi$,$\Xi_b(1P_{\lambda})\pi$,$\Lambda_b\bar{K}$ \\
	$\Xi_b$${\ket{J^P=\frac{3}{2}^-,1}}_{\rho}$&0&0&0&1&1&1&1&$\frac{3}{2}^{-}$ &&6135&6249&&6255&&$\Xi_b\pi$,$\Xi'^{(*)}_b\pi$,$\Xi_b(1P_{\lambda})\pi$,$\Lambda_b\bar{K}$ \\
    $\Xi_b$${\ket{J^P=\frac{3}{2}^-,2}}_{\rho}$&0&0&0&1&1&1&2&$\frac{3}{2}^{-}$ & &6136&6265&&6277&&$\Xi_b\pi$,$\Xi'^{(*)}_b\pi$,$\Xi_b(1P_{\lambda})\pi$,$\Lambda_b\bar{K}$&$\Xi_b(6227)?$ \\
    $\Xi_b$${\ket{J^P=\frac{5}{2}^-,2}}_{\rho}$&0&0&0&1&1&1&2&$\frac{5}{2}^{-}$ & &6133&6276&&6287& &$\Xi_b\pi$,$\Xi'^{(*)}_b\pi$,$\Xi_b(1P_{\lambda})\pi$,$\Lambda_b\bar{K}$&$\Xi_b(6227)?$\\
	\hline		
$\Xi_b^{'}$${\ket{J^P=\frac{1}{2}^-,0}}_\lambda$&0&0&1&0&1&1&0&$\frac{1}{2}^{-}$&$6238$ &$6235 $&$6201$ &6233&6198 &$6185$ &$\Xi_b\pi/\eta$, $\Xi_b'^{(*)}\pi$,$\Xi_b(1P_{\lambda})\pi$, $\Sigma_b^{(*)}K$,$\Lambda_b\bar{K}$ \\
    $\Xi_b^{'}$${\ket{J^P=\frac{1}{2}^-,1}}_\lambda$&0&0&1&0&1&1&1&$\frac{1}{2}^{-}$	&$6232$ &$6237 $&$6216$ &6227 &6220&$6220$&$\Xi_b\pi/\eta$, $\Xi_b'^{(*)}\pi$,$\Xi_b(1P_{\lambda})\pi$, $\Sigma_b^{(*)}K$,$\Lambda_b\bar{K}$  \\
	$\Xi_b^{'}$${\ket{J^P=\frac{3}{2}^-,1}}_\lambda$&0&0&1&0&1&1&1&$\frac{3}{2}^{-}$ &$6240$ &$6232 $&$6207$ &6234&6204 &$6242$&$\Xi_b\pi/\eta$, $\Xi_b'^{(*)}\pi$,$\Xi_b(1P_{\lambda})\pi$, $\Sigma_b^{(*)}K$,$\Lambda_b\bar{K}$\\
    $\Xi_b^{'}$${\ket{J^P=\frac{3}{2}^-,2}}_\lambda$&0&0&1&0&1&1&2&$\frac{3}{2}^{-}$  &$6229$ &$6234 $&$6223$ &6224&6226&$6326$ &$\Xi_b\pi/\eta$, $\Xi_b'^{(*)}\pi$,$\Xi_b(1P_{\lambda})\pi$, $\Sigma_b^{(*)}K$,$\Lambda_b\bar{K}$&$\Xi_b(6227)$\\	
	$\Xi_b^{'}$${\ket{J^P=\frac{5}{2}^-,2}}_\lambda$&0&0&1&0&1&1&2&$\frac{5}{2}^{-}$  &$6243$ &$6229 $&$6234$ &6226&6237&$6339$&$\Xi_b\pi/\eta$, $\Xi_b'^{(*)}\pi$,$\Xi_b(1P_{\lambda})\pi$, $\Sigma_b^{(*)}K$,$\Lambda_b\bar{K}$&$\Xi_b(6227)$   \\		
\hline
	$\Xi_b^{'}$${\ket{J^P=\frac{1}{2}^-,1}}_\rho$	&0&0&0&1&1&0&1&$\frac{1}{2}^{-}$&&&6366& &6367& &$\Xi_b'^{(*)}\pi$,$\Sigma_b^{(*)}K$,$\Xi_b'(1P_{\lambda})\pi$  \\
	$\Xi_b^{'}$${\ket{J^P=\frac{3}{2}^-,1}}_\rho$	&0&0&0&1&1&0&1&$\frac{3}{2}^{-}$&&&6373& &6374& & $\Xi_b'^{(*)}\pi$,$\Sigma_b^{(*)}K$,$\Xi_b'(1P_{\lambda})\pi$ \\
\hline
$\Xi_b$${\ket{J^P=\frac{1}{2}^+,0}}_\lambda$&1&0&0&0&0&0&0&$\frac{1}{2}^{+}$&$6224$ &$6208 $& &6266&6360&$6276$&$\Xi_b'^{(*)}\pi$,$\Sigma^{(*)}_bK$,$\Xi_b'(1P_{\lambda})\pi$\\	
\hline
$\Xi_b$${\ket{J^P=\frac{1}{2}^+,0}}_\rho$&0&1&0&0&0&0&0&$\frac{1}{2}^{+}$& && &&6699&&$\Xi_b'^{(*)}\pi$,$\Sigma^{(*)}_bK$,$\Xi_b'(1P_{\lambda})\pi$,$\Xi_b(1P_{\rho})\pi$,\\
&&&&&&&&&&&&&&& $\Sigma_b(1P_{\lambda})K$,$\Lambda_b(1P_{\rho})\bar{K}$,$\Xi_b'(1D_{\lambda\lambda/\rho\rho})\pi$\\		
\hline	
$\Xi_b^{'}$${\ket{J^P=\frac{1}{2}^+,1}}_\lambda$&1&0&0&0&0&1&1&$\frac{1}{2}^{+}$&$6350$&$6328$&&6329&6479 &$6367$&$\Xi_b\pi$,$\Xi_b'^{(*)}\pi$,$\Sigma^{(*)}_bK$,$\Lambda_b\bar{K}$,$\Xi_b^{(')}\eta$,$\Lambda/\Sigma B$,\\
&&&&&&&&&&&&&&& $\Xi_b(1P_{\lambda(\rho)})\pi$,$\Xi_b'(1P_{\lambda})\pi$,$\Lambda_b(1P_{\lambda})\bar{K}$\\	
$\Xi_b^{'}$${\ket{J^P=\frac{3}{2}^+,1}}_\lambda$&1&0&0&0&0&1&1&$\frac{3}{2}^{+}$	&$6370$ &$6343 $&&6342&6508&$6371$ &$\Xi_b\pi$,$\Xi_b'^{(*)}\pi$,$\Sigma^{(*)}_bK$,$\Lambda_b\bar{K}$,$\Xi_b^{(')}\eta$,$\Lambda/\Sigma B$,\\
&&&&&&&&&&&&&&& $\Xi_b(1P_{\lambda(\rho)})\pi$,$\Xi_b'(1P_{\lambda})\pi$,$\Lambda_b(1P_{\lambda})\bar{K}$\\	 
	\hline
$\Xi_b^{'}$${\ket{J^P=\frac{1}{2}^+,1}}_\rho$&0&1&0&0&0&1&1&$\frac{1}{2}^{+}$&&& &&6818 &&$\Xi_b\pi/\eta$,$\Xi_b'^{(*)}\pi/\eta$,$\Sigma^{(*)}_bK$,$\Lambda_b\bar{K}$,$\Lambda/\Sigma^{(*)} B$,\\
&&&&&&&&&&&&&&& $\Xi_b(1P_{\lambda/\rho})\pi/\eta$,$\Xi_b'(1P_{\lambda})\pi/\eta$,$\Xi_b'(1P_{\rho})\pi$,\\
&&&&&&&&&&&&&&& $\Lambda_b(1P_{\lambda/\rho})\bar{K}$,$\Sigma_b/\Omega_b(1P_{\lambda})K$,$\Sigma_b(1P_{\rho})K$,\\
&&&&&&&&&&&&&&& $\Xi^{(')}_b(1D_{\lambda\lambda/\rho\rho})\pi$,$\Lambda_b(1D_{\lambda\lambda/\rho\rho})\bar{K}$,$\Sigma_b(1D_{\lambda\lambda})K$\\
$\Xi_b^{'}$${\ket{J^P=\frac{3}{2}^+,1}}_\rho$&0&1&0&0&0&1&1&$\frac{3}{2}^{+}$	& & &&&6847&&$\Xi_b\pi/\eta$,$\Xi_b'^{(*)}\pi/\eta$,$\Sigma^{(*)}_bK$,$\Lambda_b\bar{K}$,$\Lambda/\Sigma^{(*)} B$,\\
&&&&&&&&&&&&&&& $\Xi_b(1P_{\lambda/\rho})\pi/\eta$,$\Xi_b'(1P_{\lambda})\pi/\eta$,$\Xi_b'(1P_{\rho})\pi$,\\
&&&&&&&&&&&&&&& $\Lambda_b(1P_{\lambda/\rho})\bar{K}$,$\Sigma_b/\Omega_b(1P_{\lambda})K$,$\Sigma_b(1P_{\rho})K$,\\
&&&&&&&&&&&&&&& $\Xi^{(')}_b(1D_{\lambda\lambda/\rho\rho})\pi$,$\Lambda_b(1D_{\lambda\lambda/\rho\rho})\bar{K}$,$\Sigma_b(1D_{\lambda\lambda})K$\\ 	\hline
\bottomrule[1.2pt]\end{tabular}  %
}
\end{table*}

This paper is structured as follows. In Sec. II, we briefly
introduce the QPC model, chiral quark model and the relationship of states in different coupling schemes.
Then we present our numerical results and discussions in Sec. III. A summary is given in Sec. IV.

\section{Theoretical framework}

\subsection{QPC model}\label{model}

The QPC model was first proposed by Micu \cite{Micu:1968mk}, Car-litz and Kislinger \cite{R1970Regge}, and further developed by the Orsay group \cite{LeYaouanc:1972vsx,LeYaouanc:1977fsz,LeYaovanc1988Hadron}.
According to this model, the strong decay takes place via the creation of quark-antiquark pair from the vacuum with quantum number $0^{++}$.
For a baryon decay process, one created quark regroups with two of the initial baryon state $|A\rangle$ to
form a daughter baryon $|B\rangle$, and the created antiquark regroups with the other one quark to form a meson $|C\rangle$.

%Hence, there are three possible decay ways for the $\Xi^{(')}_b$ resonances, as shown in Fig~\ref{FIG1}.

Within the QPC model, the transition operator for the two-body decay process $A\to BC$ in the nonrelativistic limit is given by
\begin{equation}
	\begin{aligned}
	T=&-3\gamma\sum\limits_{m}\langle1m;1-m|00\rangle\int d^3\textbf{p}_4d^3\textbf{p}_5\delta^3(\textbf{p}_4+\textbf{p}_5)\\ &\times\mathcal{Y}^m_1(\dfrac{\textbf{p}_4-\textbf{p}_5}{2})\chi^{45}_{1-m}\phi^{45}_0\omega^{45}_0a^\mathcal{y}_{4i}(\textbf{p}_4)b^\mathcal{y}_{5j}(\textbf{p}_5).\\
\end{aligned}
\end{equation}
The dimensionless parameter $\gamma$ accounts for the vacuum pair-production strength.
$\textbf{p}_4$ and $\textbf{p}_5$ are the three-vector momenta of the created quark pair. $\mathcal{Y}^m_1$=$|\textbf{p}|$$\rm{\boldsymbol{Y}}^m_1$($\theta_p\phi_p$) is solid harmonic polynomial and reflects the momentum-space distribution.
$\chi^{45}_{1-m}$, $\phi^{45}_0$=(u$\rm{\overline{u}}$+d$\rm{\overline{d}}$+s$\rm{\overline{s}}$)/$\sqrt{3}$, and $\omega^{45}_0$=$\delta_{ij}$ are the spin triplet, flavor function, and color singlet, respectively.
$a^\mathcal{y}_{4i}d^\mathcal{y}_{5j}$ is the creation operator and denotes the quark pair-creation in the vacuum.

 Describing the spatial wave functions of baryons and mesons with simple harmonic oscillator wave functions in momentum space, then we can obtain the partial decay amplitude in the center of mass frame,
\begin{equation}
	\begin{aligned}
		&\mathcal{M}^{M_{J_A}M_{J_B}M_{J_C}}(A\rightarrow BC)\\
		=&\gamma\sqrt{8E_AE_BE_C}\prod_{A,B,C}\langle\chi^{124}_{S_BM_{S_B}}\chi^{35}_{S_CM_{S_C}}|\chi^{123}_{S_AM_{S_A}}\chi^{45}_{1-m}\rangle\\
		&\langle\varphi^{124}_B\varphi^{35}_C|\varphi^{123}_A\varphi^{45}_0\rangle I^{M_{L_A},m}_{M_{L_B},M_{L_C}}(\textbf{p}).
	\end{aligned}
\end{equation}
Here, $I^{M_{L_A},m}_{M_{L_B},M_{L_C}}(\textbf{p})$ is the spatial integration. $\prod_{A,B,C}$ denotes the Clebsch-Gorden coefficients and reads
\begin{equation}
	\begin{aligned}
			\sum&\langle L_BM_{L_B};S_BM_{S_B}|J_B,M_{J_B}\rangle \langle L_CM_{L_C};S_CM_{S_C}|J_C,M_{J_C}\rangle\\
			 &\times \langle L_AM_{L_A};S_AM_{S_A}|J_A,M_{J_A}\rangle \langle1m;1-m|00\rangle.\\
	\end{aligned}
\end{equation}
It should be mentioned that the vertices described by the transition operator $T$ are only effective in the non-perturbative region, which characters the capability of creating a quark-antiquark pair from the vacuum. This capability will be suppressed in the high momentum region due to the weakening of valence quark interactions. To suppress the nonphysical contributions in high momentum region due to the overly hard vertices given by the quark pair creation model, we introduce a suppressed form factor $e^{-\frac{\mathbf{p}^2}{2\Lambda^2}}$ into the decay amplitude,
\begin{equation}
	\begin{aligned}
		\mathcal{M}^{M_{J_A}M_{J_B}M_{J_C}}(A\rightarrow BC)\rightarrow \mathcal{M}^{M_{J_A}M_{J_B}M_{J_C}}(A\rightarrow BC)e^{-\frac{\mathbf{p}^2}{2\Lambda^2}}.
	\end{aligned}
\end{equation}
as that done in the literature~\cite{Zhang:2024afw,Ortega:2016mms,Morel:2002vk,Ortega:2016pgg}.
In the equation, we fix the cut-off parameter $\Lambda=780$ MeV, the same value as used in Ref.~\cite{Ni:2023lvx}. $\textbf{p}$ denotes the momentum of the daughter baryon $|B\rangle$ in the center of mass frame of baryon $|A\rangle$, its calculation is given by the following formula:
\begin{equation}
	|\textbf{p}|=\dfrac{\sqrt{[M_A^2-(M_B-M_C)^2][M_A^2-(M_B+M_C)^2]}}{2M_A}.
\end{equation}

Finally, the decay width of the process $A\to BC$ can be obtained by the following formula,
\begin{equation}
\Gamma(A\rightarrow BC)=\pi^2 \dfrac{|\textbf{p}|}{M^2_A}\dfrac{1}{2J_A+1}\sum_{M_{J_A},M_{J_B},M_{J_C}}|\mathcal{M}^{M_{J_A},M_{J_B},M_{J_C}}e^{-\frac{\mathbf{p}^2}{2\Lambda^2}}|^2.
\end{equation}

In this work, we adopt $m_u=m_d=330$ MeV, $M_s=450$ MeV, and $m_b=5000$ MeV for the constituent quark mass.
The other parameters, such as harmonic oscillator strength and vacuum pair-production strength, are the same as in our previous article~\cite{Zhou:2023xxs}.

\subsection{Chiral quark model}

In the framework of the chiral quark model (ChQM), the low energy effective quark-pseudoscalar-meson coupling at tree level in the SU(3) flavor basis reads
~\cite{Manohar:1983md}
\begin{eqnarray}\label{STcoup}
\mathcal{L}_m=\sum_j
\frac{\delta}{\sqrt{2}f_m}\bar{\psi}_j\gamma^{j}_{\mu}\gamma^{j}_{5}\psi_j\vec{I}\cdot\partial^{\mu}\vec{\phi}_m.
\end{eqnarray}
In the equation, $f_m$ denotes the pseudoscalar meson decay constant, and $\delta$ is a global parameter accounting for the strength of the quark-pseudoscalar-meson couplings.
$\psi_j$ represents the $j$-th quark field in a baryon. $I$ and $\phi_m$ denote an isospin operator and the pseudoscalar meson octet, respectively.
%which is
% \begin{eqnarray}
%\phi_m=
%\begin{pmatrix}
%\frac{1}{\sqrt{2}}\pi^0+\frac{1}{\sqrt{6}}\eta & \pi^+ & K^+ \cr
%\pi^- & -\frac{1}{\sqrt{2}}\pi^0+\frac{1}{\sqrt{6}}\eta & K^0 \cr
%K^- & \bar{K}^0 & -\sqrt{\frac{2}{3}\eta}
%\end{pmatrix}.
%\end{eqnarray}
 Considering the harmonic oscillator spatial wave function of baryons being nonrelativistic form, the coupling is adopt the nonrelativistic form as well and is written as~\cite{Zhong:2024mnt,Ni:2023lvx,Zhong:2025oti}
\begin{eqnarray}\label{non-relativistic-expansST}
H_{I}=\mathcal{H}^{NR}+\mathcal{H}^{RC},
\end{eqnarray}
with
\begin{eqnarray}
\mathcal{H}^{NR}=g\sum_j\left(\mathcal{G}\vsig_j\cdot\textbf{q}+ \frac{\omega_m}{2\mu_q}\vsig_j \cdot\textbf{p}_j\right)I_j \varphi_m,
\end{eqnarray}
and
\begin{equation}
\begin{aligned}
 \mathcal{H}^{RC}=&-\frac{g}{32\mu_q^2}\sum_j\bigg[m^2(\vsig_j\cdot\textbf{q})\\
 &+2\vsig_j\cdot(\textbf{q}-2\textbf{p}_j)\times(\textbf{q}\times\textbf{p}_j)\bigg]I_j\varphi_m.
\end{aligned}
\end{equation}
In the above equation, $\vsig_j$ and $\textbf{p}_j$ are the spin operator and internal momentum operator of the $j$-th light quark within a hadron. $\textbf{q}$, $\omega_m$, $\varphi_m$ and $m$ are the three momentum, energy, plane wave, and mass of the final emitted pseudoscalar light meson, respectively. The reduced mass $\mu_q$ is expressed as $1/\mu_q=1/m_j+1/m'_j$ for the masses of the $j$-th quark in the initial and final baryons. $I_j$ is the isospin operator. $g$ and $\mathcal{G}$ are defined by $g=\delta\sqrt{(E_i+M_i)(E_f+M_f)}$ and $\mathcal{G}=-(\frac{\omega_m}{E_f+M_f}+1+\frac{\omega_m}{2m'_j})$, respectively. Here, $(E_{i/f}, M_{i/f})$ are the energy and mass of the initial/final baryon.

Within the ChQM, the strong decay amplitude for the $A$(initial baryon)$\rightarrow$ $B$(final baryon)+$C$(final meson) process can be obtained with
\begin{equation}
\mathcal{M}(A\rightarrow BC)=\langle B|H_I|A\rangle.
\end{equation}
Similarly, the vertices described by the $H_I$ are only effective in the non-perturbative region. To suppress the unphysical contributions in high momentum region, we introduce the same form factor $e^{-\frac{\mathbf{p}^2}{2\Lambda^2}}$ into the decay amplitude,
\begin{equation}
	\begin{aligned}
		\mathcal{M}_{J_{iz},J_{fz}}\rightarrow \mathcal{M}_{J_{iz},J_{fz}}e^{-\frac{\mathbf{p}^2}{2\Lambda^2}}.
	\end{aligned}
\end{equation}

Then, the partial decay width can be worked out by
\begin{equation}
\Gamma=\frac{1}{8\pi}\frac{|\mathbf{q}|}{M_i^2}\frac{1}{2J_i+1}\sum_{J_{iz},J_{fz}}|\mathcal{M}_{J_{iz},J_{fz}}e^{-\frac{\mathbf{p}^2}{2\Lambda^2}}|^2.
\end{equation}
Here, $J_i$ is the total angular momentum quantum number of the initial baryon. $J_{iz}$ and $J_{fz}$ are the third components of the total angular momenta of the initial and final baryons, respectively.

In the calculations, the constituent quark mass, harmonic oscillator strength, and cut-off parameter are the same as in that of the QPC model. The other model parameters are the same as in Refs.~\cite{Zhong:2024mnt,Ni:2023lvx,Zhong:2025oti}. It should be mentioned that the strong decay properties of the excited $\Xi_b$ and $\Xi'_b$ baryons have been studied within the QPC model~\cite{Chen:2018orb,Bijker:2020tns,Garcia-Tecocoatzi:2023btk,He:2021xrh} and ChQM~\cite{Wang:2017kfr,Wang:2018fjm}. However, in these works, both the QPC model and ChQM does not account for high-momentum corrections. Meanwhile, ChQM neglects relativistic correction term $\mathcal{H}^{RC}$. To enhance the reliability of the results, we have incorporated the aforementioned corrections in this paper.

\subsection{Coupling scheme}

Due to the heavy quark symmetry, the physical states for the $\Xi_b$ and $\Xi_b'$ baryons may be closer to the $j$-$j$ coupling scheme.
In the heavy quark symmetry limit, the states within the $j$-$j$ coupling scheme are constructed by
\begin{equation}
|J^P,j\rangle=	\left|\left\{[(l_{\rho}l_{\lambda})_L s_{\rho}]_j s_Q\right\}_{J^P}
	\right \rangle.
\end{equation}
In the expression, $l_{\rho}$ and $l_{\lambda}$ are the quantum numbers of the orbital angular momentum for
$\rho$- and $\lambda$-mode excitations, respectively. The the quantum number of total orbital angular momentum
$L = |l_{\rho}-l_{\lambda}|,\cdots , l_{\rho}+l_{\lambda} $. $s_{\rho}$ is the spin quantum number of
the light quark pair, and $s_Q$ is the spin quantum number of the
heavy quark.

Meanwhile, the states within the the
$j$-$j$ coupling scheme can be expressed as linear combinations
of the states within the $L$-$S$ coupling via~\cite{Roberts:2007ni}
\begin{equation}
\begin{aligned}
	\left|\left\{[(l_{\rho}l_{\lambda})_L s_{\rho}]_j s_Q\right\}_{J^P}
	\right \rangle&=(-1)^{L+s_{\rho}+\frac{1}{2}+J}\sqrt{2J+1}\sum_{S}\sqrt{2S+1}\\
	&\begin{pmatrix}
		L&s_{\rho}&j\\s_Q&J&S
	\end{pmatrix}	\left|\left\{[(l_{\rho}l_{\lambda})_L (s_{\rho}s_Q)_S]_J \right\}
\right\rangle.
\end{aligned}
\end{equation}
The quantum number of total spin angular momentum $S =|s_{\rho}-s_Q|,\cdots , s_{\rho}+s_Q $.
$J$ is the quantum number of total angular momentum.

\section{Calculations and Results }

We calculate the strong decays of
 $1P$- and $2S$-wave excited $\Xi_b$ and $\Xi'_b$ baryons within the $j$-$j$ coupling scheme
in the framework of the QPC model. Both $\lambda$-mode and $\rho$-mode excitations are considered in the present work.
In addition, for a comparison we also give the decay properties of those $\Xi_b$ and $\Xi'_b$ excitations within the ChQM.
It is anticipated that our theoretical calculations will not only shed light on the internal structure of the presently observed controversial states, but also provide a theoretical foundation for future experimental investigations of the "missing" resonances.

%Here, we mainly focus on the theoretical results within the QPC model and just simply present the calculations obtained by the chiral quark model.

\subsection{$1P$-wave $\lambda$-mode excitations}

\begin{figure}[]
	\centering \epsfxsize=8.5 cm \epsfbox{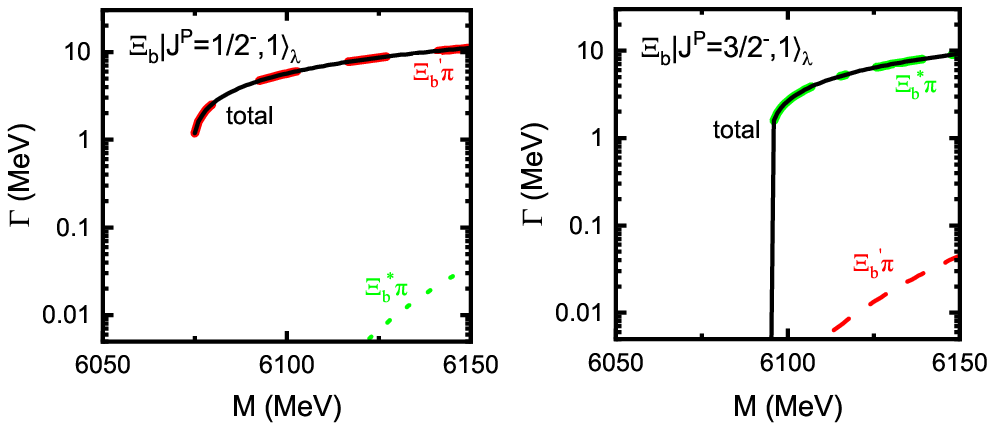}
	\caption{ Partial and total strong decay widths of the $1P$-wave $\lambda$-mode $\Xi_b$ excitations as functions of their masses.}
    \label{FIG1}
\end{figure}

For the $1P$-wave $\lambda$-mode $\Xi_b$ baryons, there are two states according to the quark model classification, which are $\Xi_b|J^P=1/2^-,1\rangle_{\lambda}$ and $\Xi_b|J^P=3/2^-,1\rangle_{\lambda}$. Their theoretical masses and possible two-body decay channels are listed in Table~\ref{table2}. From the table, it is known that the predicted masses of the $1P$-wave $\lambda$-mode $\Xi_b$ baryons are about $M\simeq(6050-6150)$ MeV, and the possible two-body decay channels are $\Xi_b'^{(*)}\pi$. Considering the uncertainty of the mass predictions, we plot the strong decay properties of the $\lambda$-mode $1P$-wave $\Xi_b$ states as a function of the mass in Fig.~\ref{FIG1} with the QPC model.

\begin{table}[]
	\caption{\label{Xiplambda}The strong decay properties of the $\lambda$-mode $1P$-wave $\Xi_b$ states, which are taken as $\Xi_b(6087)^0$ and $\Xi_b(6095)^0$, respectively. The symbol ``-'' indicates that this decay channel is forbidden. $\Gamma_{\text{Total}}$ represents the total decay width, and Expt. denotes the experimental value. The units are MeV.}
	\centering
	\begin{tabular}{ccccccccccccccccc}	
		\hline\hline		&\multicolumn{2}{c}{$\underline{~\Xi_{b}\ket{J^{P}=\frac{1}{2}^{-},1}_{\lambda}~}$}&\multicolumn{2}{c}{$\underline{~\Xi_{b}\ket{J^{P}=\frac{3}{2}^{-},1}_{\lambda}~}$}\\
 &\multicolumn{2}{c}{$\Xi_b(6087)^0$}&\multicolumn{2}{c}{$\Xi_b(6095)^0$}\\
				\cline{2-5}
Decay width~~&QPC&ChQM&QPC&ChQM\\
		\hline				
		$\Gamma[\Xi_b'^-\pi^+]$&1.2&0.8&0.0&0.0\\
		$\Gamma[\Xi_b'^0\pi^0]$&1.4&0.5&0.0&0.0\\
        $\Gamma[\Xi_b^{*-}\pi^+]$&-&-&0.1&0.1\\
		$\Gamma[\Xi_b^{*0}\pi^0]$&-&-&0.9&0.3\\
\hline
        $\Gamma_{\text{Total}}$&2.6&1.3&1.0&0.4\\
        \hline
	Expt.&\multicolumn{2}{c}{$2.4\pm0.5$}&\multicolumn{2}{c}{$0.50\pm0.35$}\\	
\hline \hline
\end{tabular}	
\end{table}

It is obtained that the decay properties of $\Xi_b|J^P=1/2^-,1\rangle_{\lambda}$ and $\Xi_b|J^P=3/2^-,1\rangle_{\lambda}$ are sensitive to the masses. Meanwhile, the two $1P$-wave $\lambda$-mode $\Xi_b$ baryons are probably very narrow states with a comparable total decay width of a few MeV. However, their main decay channels are completely different. The total decay width of $\Xi_b|J^P=1/2^-,1\rangle_{\lambda}$ is almost saturated by $\Xi_b'\pi$, while that of $\Xi_b|J^P=3/2^-,1\rangle_{\lambda}$ is almost saturated by $\Xi_b^{*}\pi$. Combining the measured masses and decay properties of the observed states $\Xi_b(6087)^0$ and $\Xi_b(6095)^0$, which mainly decay into $\Xi_b'\pi$ and $\Xi_b^{*}\pi$, respectively, $\Xi_b|J^P=1/2^-,1\rangle_{\lambda}$ and $\Xi_b|J^P=3/2^-,1\rangle_{\lambda}$ are good candidates of the two observed states. Hence, we further take $\Xi_b|J^P=1/2^-,1\rangle_{\lambda}$ and $\Xi_b|J^P=3/2^-,1\rangle_{\lambda}$ as $\Xi_b(6087)^0$ and $\Xi_b(6095)^0$, respectively, and collect their decay properties in Table~\ref{Xiplambda}. For a comparison, we also studied the decay properties within the ChQM and listed the results in the same table as well.

Within the QPC model, the total decay width of $\Xi_b|J^P=1/2^-,1\rangle_{\lambda}$ is
\begin{equation}
\Gamma_{\mathrm{Total}}\simeq2.6 \mathrm{MeV},
\end{equation}
which is twice as large as that of the ChQM and highly consistent with the observation. Meanwhile, the dominant decay channels are $\Xi_b'^-\pi^+$ and $\Xi_b'^0\pi^0$. The corresponding branching fractions with the QPC model are
\begin{equation}
\frac{\Gamma[\Xi_b|J^P=1/2^-,1\rangle_{\lambda}\rightarrow \Xi_b'^-\pi^+]}{\Gamma_{\mathrm{Total}}}\simeq46\%
\end{equation}
and
\begin{equation}
\frac{\Gamma[\Xi_b|J^P=1/2^-,1\rangle_{\lambda}\rightarrow \Xi_b'^0\pi^0]}{\Gamma_{\mathrm{Total}}}\simeq54\%,
\end{equation}
respectively. This calculation is consistent with the fact that $\Xi_b(6087)^0$ was observed in $\Xi_b^0\pi^-\pi^+$ final state mainly via the intermediate channel $\Xi_b'^-\pi^+$ by LHCb Collaboration~\cite{LHCb:2023zpu}. In addition, if $\Xi_b(6087)^0$ corresponds to the state $\Xi_b|J^P=1/2^-,1\rangle_{\lambda}$ indeed, $\Xi_b(6087)^0\rightarrow \Xi_b'^0\pi^0\rightarrow \Xi_b^0\pi^0\pi^0$ may be another interesting decay chain for experimental exploration for the large predicted branching ratio of the $\Xi_b'^0\pi^0$ channel. However, this neutral decay channel will pose a challenge for experimental detection.

Our interpretations concur with prevailing theory in suggestion state $\Xi_b(6087)^0$ being a spin-parity $J^P=1/2^-$ state, which directly contradicts the experimental value of $J^P=3/2^-$. This discrepancy necessitates a careful experimental reassessment of the $\Xi_b(6087)^0$ state's spin-parity. It should mentioned that although its spin-parity is consistently interpreted as $J^P=1/2^-$ across different theoretical approaches, the classification of this state within the quark model remains diverse. Based the mass spectrum~\cite{Bijker:2023zhh,Li:2023gbo,Xin:2023usd}, the authors interpreted $\Xi_b(6087)^0$ as $1P$ $\lambda$-mode $\Xi_b$ state. While, based on both the mass spectrum and strong decay properties~\cite{Wang:2024rai,Tan:2023opd}, the authors interpreted $\Xi_b(6087)^0$ as $1P$ $\rho$-mode $\Xi_b$ state.

For the state $\Xi_b|J^P=3/2^-,1\rangle_{\lambda}$, the total decay width within the QPC model is
\begin{equation}
\Gamma_{\mathrm{Total}}\simeq1.0 ~\mathrm{MeV},
\end{equation}
which is close to the upper limit of the experimental value and approximately twice that of the ChQM. The main decay channels are $\Xi_b^{*-}\pi^+$ and $\Xi_b^{*0}\pi^0$, and the corresponding branching ratios in the QPC model are
\begin{equation}
\frac{\Gamma[\Xi_b|J^P=3/2^-,1\rangle_{\lambda}\rightarrow \Xi_b^{*-}\pi^+]}{\Gamma_{\mathrm{Total}}}\simeq10\%
\end{equation}
and
\begin{equation}
\frac{\Gamma[\Xi_b|J^P=1/2^-,1\rangle_{\lambda}\rightarrow \Xi_b^{*0}\pi^0]}{\Gamma_{\mathrm{Total}}}\simeq90\%.
\end{equation}
Considering the uncertainties for the experimental data and theoretical calculations, the decay properties of $\Xi_b|J^P=3/2^-,1\rangle_{\lambda}$ are consistent with those of $\Xi_b(6095)^0$. Additionally, due to the large branching ratio of decay channel $\Xi_b^{*0}\pi^0$, the $\Xi_b(6095)^0$ state are likely to be observed in the $\Xi_b^-\pi^+\pi^0$ final state via the decay chain $\Xi_b(6095)^0\rightarrow \Xi_b^{*0}\pi^0\rightarrow \Xi_b^-\pi^+\pi^0$, apart from the already experimentally observed decay channel $\Xi_b^0\pi^-\pi^+$. It should be mentioned that the mass of $\Xi_b(6095)^0$ is very close to the threshold of $\Xi_b^*\pi$, and the partial decay widths are highly sensitive to the precision of mass.

\begin{figure}[]
	\centering \epsfxsize=8.5 cm \epsfbox{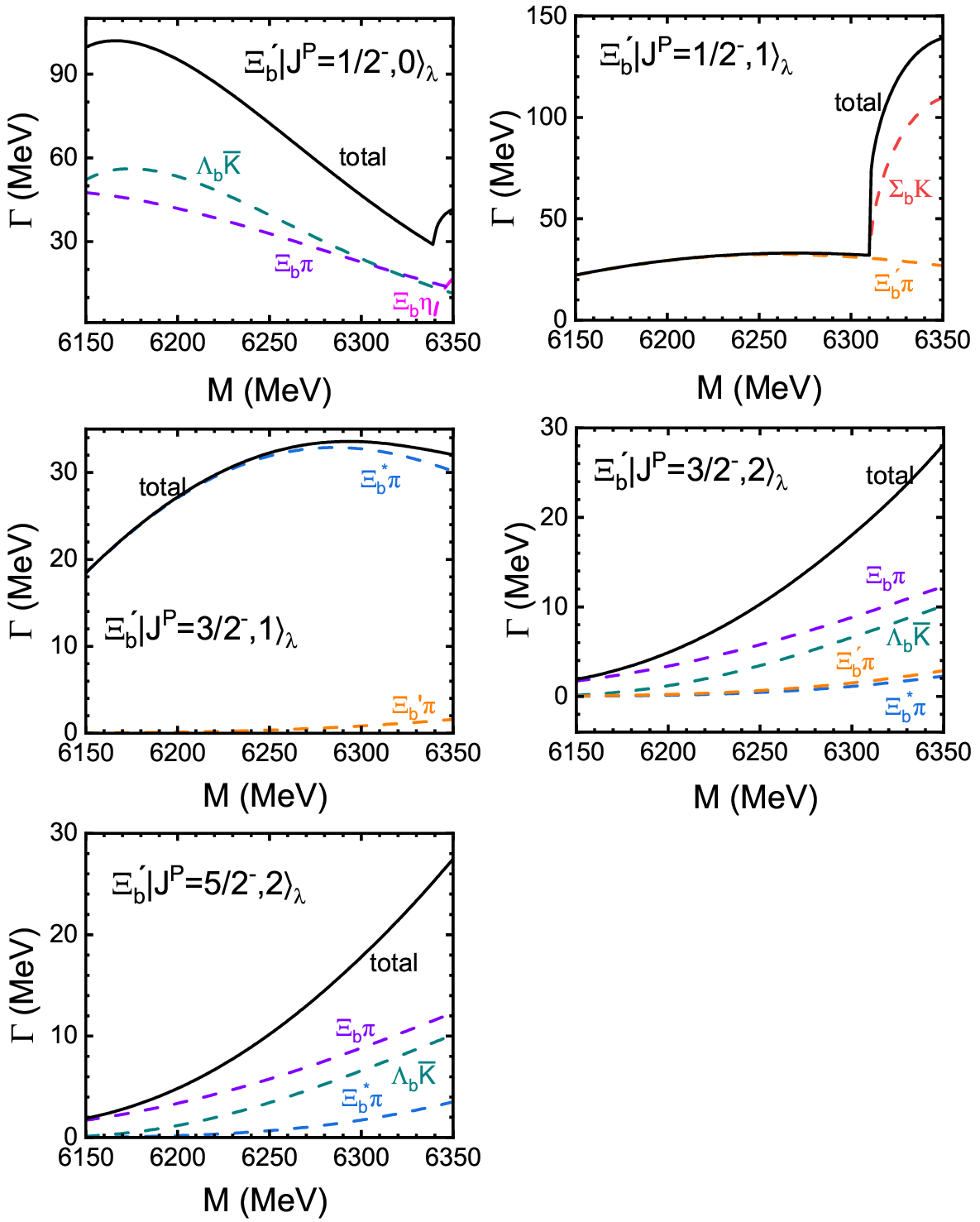}
	\caption{ Partial and total strong decay widths of the $1P$-wave $\lambda$-mode $\Xi_b'$ excitations as functions of their masses. Some decay channels are too small to show in figure.}
    \label{FIG2}
\end{figure}

As to the $1P$-wave $\lambda$-mode $\Xi_b'$ baryons, there are five states $\Xi_b'|J^P=1/2^-,0\rangle_{\lambda}$, $\Xi_b'|J^P=1/2^-,1\rangle_{\lambda}$, $\Xi_b'|J^P=3/2^-,1\rangle_{\lambda}$, $\Xi_b'|J^P=3/2^-,2\rangle_{\lambda}$, and $\Xi_b'|J^P=5/2^-,2\rangle_{\lambda}$. According to the theoretical predictions by various methods, the mass of the $1P$-wave $\lambda$-mode $\Xi_b'$ baryons is about $M\sim6.23$ GeV(see table~\ref{table2}). Based on the predicted masses, these states are probably assignments of observed state $\Xi_b(6227)$. Hence, the investigation of the properties of these states is crucial for clarifying the internal structure of the state $\Xi_b(6227)$. Considering the uncertainty of the mass predictions as well, we plot the decay properties of the $1P$-wave $\lambda$-mode $\Xi_b'$ baryons as functions of masses within the range of $M=(6150-6350)$ MeV within the QPC model, as show in Fig.~\ref{FIG2}.

\begin{table*}[]
	\caption{\label{Xipplambda}The strong decay properties of the $\lambda$-mode $1P$-wave $\Xi_b'$ states, which masses are taken from the predictions in Ref.~\cite{Ortiz-Pacheco:2023kjn}. The symbol ``-'' indicates that this decay channel is forbidden. $\Gamma_{\text{Total}}$ represents the total decay width, and Expt. denotes the experimental value. The units are MeV.}
	\centering
	\begin{tabular}{ccccccccccccccccc}	
		\hline\hline		&\multicolumn{2}{c}{$\underline{~\Xi_{b}'\ket{J^{P}=\frac{1}{2}^{-},0}_{\lambda}~}$}&\multicolumn{2}{c}{$\underline{~\Xi_{b}'\ket{J^{P}=\frac{1}{2}^{-},1}_{\lambda}~}$}
&\multicolumn{2}{c}{$\underline{~\Xi_{b}'\ket{J^{P}=\frac{3}{2}^{-},1}_{\lambda}~}$}&\multicolumn{2}{c}{$\underline{~\Xi_{b}'\ket{J^{P}=\frac{3}{2}^{-},2}_{\lambda}~}$}
&\multicolumn{2}{c}{$\underline{~\Xi_{b}'\ket{J^{P}=\frac{5}{2}^{-},2}_{\lambda}~}$}\\
 &\multicolumn{2}{c}{$M=6201$}&\multicolumn{2}{c}{$M=6216$}&\multicolumn{2}{c}{$M=6207$}&\multicolumn{2}{c}{$\Xi_b(6227)^0$}&\multicolumn{2}{c}{$\Xi_b(6227)^0$}\\
				\cline{2-11}
Decay width~~&QPC&ChQM&QPC&ChQM&QPC&ChQM&QPC&ChQM&QPC&ChQM\\
		\hline				
		$\Gamma[\Xi_b\pi]$&41.8&10.3&0.0&0.0&0.0&0.0&4.6&14.2&4.6&14.2\\
		$\Gamma[\Xi_b'\pi]$&0.0&0.0&30.7&12.5&0.1&0.4&0.4&1.1&0.2&0.5\\
        $\Gamma[\Xi_b^{*}\pi]$&0.0&0.0&0.2&0.6&28.0&11.8&0.3&0.7&0.4&1.3\\
		$\Gamma[\Lambda_b\bar{K}]$&53.2&13.4&0.0&0.0&0.0&0.0&2.3&7.4&2.3&7.4\\
$\Xi_b\ket{J^{P}=\frac{1}{2}^{-},1}_{\lambda}\pi$&-&-&-&-&-&-&0.0&0.0&0.0&0.0\\
\hline
        $\Gamma_{\text{Total}}$&95.0&23.7&30.9&13.1&28.1&12.2&7.6&23.4&7.5&23.4\\
        \hline
	Expt.&\multicolumn{2}{c}{}&\multicolumn{2}{c}{}&\multicolumn{2}{c}{}&\multicolumn{4}{c}{$19^{+5}_{-4}$}\\	
\hline \hline
\end{tabular}	
\end{table*}

We can find that $\Xi_b'|J^P=1/2^-,0\rangle_{\lambda}$ is likely a moderate state with a decay width on the order of tens of MeV. The dominant decay channels are $\Lambda_b\bar{K}$ and $\Xi_b\pi$. For $\Xi_b'|J^P=1/2^-,1\rangle_{\lambda}$, its total decay width is about 20-30 MeV, and the dominant decay channel is $\Xi_b'\pi$. However, we notice that if its mass lies above the $\Sigma_bK$ threshold, this mode becomes dominant and the total decay width can increase dramatically to well over 100 MeV. The total decay width of $\Xi_b'|J^P=3/2^-,1\rangle_{\lambda}$ is comparable to that of $\Xi_b'|J^P=1/2^-,1\rangle_{\lambda}$, both on the order of 20-30 MeV. While $\Xi_b'|J^P=3/2^-,1\rangle_{\lambda}$ mainly decays into $\Xi^*\pi$, which stands in sharp contrast to that of $\Xi_b'|J^P=1/2^-,1\rangle_{\lambda}$. Considering the relatively narrow width, $\Xi_b'|J^P=3/2^-,1\rangle_{\lambda}$ is highly likely to be observed in the $\Xi_b\pi\pi$ final state via the decay chain $\Xi_b'|J^P=3/2^-,1\rangle_{\lambda}\rightarrow \Xi^*\pi\rightarrow \Xi_b\pi\pi$. $\Xi_b'|J^P=3/2^-,2\rangle_{\lambda}$ and $\Xi_b'|J^P=5/2^-,2\rangle_{\lambda}$ share similar decay properties. Specifically, both have total decay widths less than 30MeV, and their dominant decay modes are $\Xi_b\pi$ and $\Lambda_b\bar{K}$. Based on the decay properties of the observed state $\Xi_b(6227)$, we cannot rule out the possibility that $\Xi_b(6227)$ could be a candidate of $\Xi_b'|J^P=3/2^-,2\rangle_{\lambda}$ or $\Xi_b'|J^P=5/2^-,2\rangle_{\lambda}$.

To better visualize the decay properties of the $1P$-wave $\lambda$-mode $\Xi_b'$ baryons, we further calculate their decay properties with the masses fixed at the predictions in Ref.~\cite{Ortiz-Pacheco:2023kjn} within both the QPC model and ChQM, except for the states $\Xi_b'|J^P=3/2^-,2\rangle_{\lambda}$ and $\Xi_b'|J^P=5/2^-,2\rangle_{\lambda}$, which masses are fixed at the physical mass of $\Xi_b(6227)^0$. The results are listed in Table~\ref{Xipplambda}.

According to the results of the QPC model, when $\Xi_b'|J^P=3/2^-,2\rangle_{\lambda}$ and $\Xi_b'|J^P=5/2^-,2\rangle_{\lambda}$ are treated as $\Xi_b(6227)^0$, both have a total width of about
\begin{equation}
\Gamma_{\mathrm{Total}}\simeq7.6-7.5~ \mathrm{MeV},
\end{equation}
 which is approximately (40-39)\% of the experimental central value. While, within the ChQM, the total decay widths of the two states are
 \begin{equation}
\Gamma_{\mathrm{Total}}\simeq23.4~ \mathrm{MeV},
\end{equation}
 which is in good agreement with the experimental result, within the experimental uncertainty.
 Additionally, both predominantly decay to $\Xi_b\pi$(including $\Xi_b^0\pi^0$ and $\Xi_b^-\pi^+$) and $\Lambda_b\bar{K}$, and the branching fraction ratio in the QPC and ChQM is
\begin{equation}
\frac{\Gamma[\Xi_b'|J^P=3(5)/2^-,2\rangle_{\lambda}\rightarrow \Lambda_b\bar{K}]}{\Gamma[\Xi_b'|J^P=3(5)/2^-,2\rangle_{\lambda}\rightarrow \Xi_b\pi]}\simeq0.5.
\end{equation}
This value is highly consistent with the experimental measurement ($B(\Xi_b(6227)^-\rightarrow\Lambda_bK)/B(\Xi_b(6227)^-\rightarrow\Xi_b^0\pi^-)\simeq1.0\pm0.5$) by the LHCb Collaboration~\cite{LHCb:2018vuc}. Hence, $\Xi_b'|J^P=3/2^-,2\rangle_{\lambda}$ and $\Xi_b'|J^P=5/2^-,2\rangle_{\lambda}$ could be good candidate states for the observed state $\Xi_b(6227)^0$.

\begin{figure}[]
	\centering \epsfxsize=6 cm \epsfbox{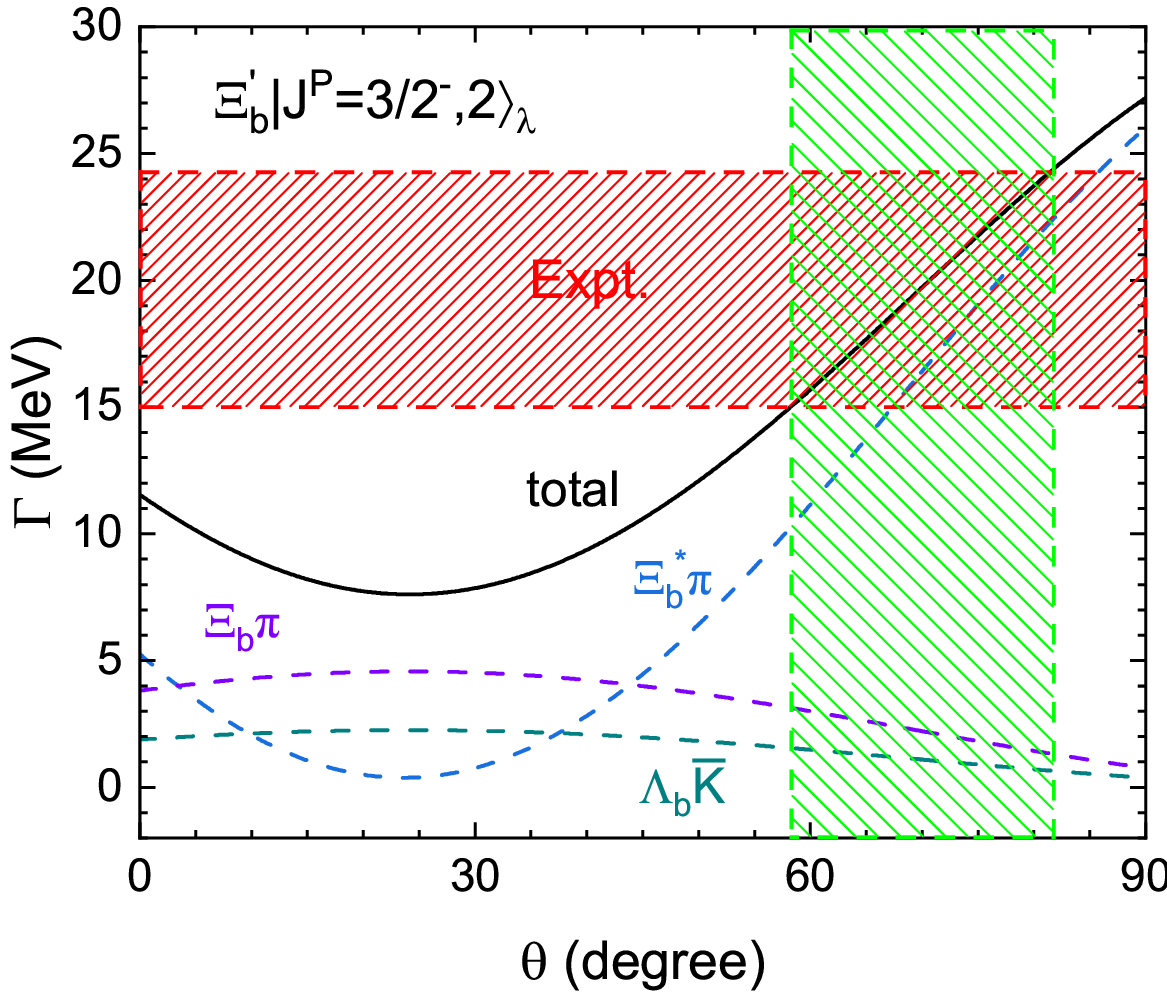}
	\caption{ Partial and total strong decay widths of $\Xi_b'|J^P=3/2^-,2\rangle_{\lambda}$ as a function of the mixing angle. Some decay channels are too small to show in figure.}
    \label{FIG3}
\end{figure}

Additionally, we know that the state $\Xi_b'|J^P=3/2^-,2\rangle_{\lambda}$ in the $j-j$ coupling scheme is the linear combination of the configurations in the $L-S$ coupling scheme. That means $\Xi_b'|J^P=3/2^-,2\rangle_{\lambda}$ in the $j-j$ coupling scheme contains a mixing angle $\theta$. Considering the heavy quark symmetry being not strictly true and slightly breaking in the $\Xi_b'$ system, the mixing angle $\theta$ will fluctuate around the center value($\theta\simeq24^{\circ}$). To investigate this effect, we plot the strong decay properties of $\Xi_b'|J^P=3/2^-,2\rangle_{\lambda}$ within the QPC model as a function of the mixing angle varying in region of $\theta=(0^{\circ}-90^{\circ})$ in Fig.~\ref{FIG3}.
As shown in the figure, when the mixing angle $\theta$ ranges from $58^{\circ}$ to $81^{\circ}$, the total decay width of $\Xi_b'|J^P=3/2^-,2\rangle_{\lambda}$ is between 15 and 24 MeV, which is consistent with that of the observed state $\Xi_b(6227)^0$. Furthermore, considering the fact that $\Xi_b(6227)^0$ is observed in $\Xi_b^-\pi^+$~\cite{LHCb:2020xpu} and the experimental measured partial width ratio~\cite{LHCb:2018vuc} of the decay channel $\Lambda_b\bar{K}$ to $\Xi_b\pi$ is approximately equal to one, selecting the lower limit of the allowed range for the mixing angle($\theta\simeq58^{\circ}$) is more appropriate. Under this condition, the branching ratios of decay channels $\Lambda_b\bar{K}$ and $\Xi_b\pi$ are approximately 11\% and 21\% respectively. Additionally, the decay channel $\Xi_b^*\pi$ will occupy a considerably large branching ratio of about 68\%. Hence $\Xi_b\pi\pi$ could be another ideal observable channel for the state $\Xi_b(6227)^0$ via the decay chain $\Xi_b(6227)\rightarrow \Xi_b^*\pi\rightarrow \Xi_b\pi\pi$.

\subsection{$1P$-wave $\rho$-mode excitations}

According to the symmetry of wave functions, there are five $\rho$-mode $1P$-wave $\Xi_b$ baryons(see Table~\ref{table2}): $\Xi_b|J^P=1/2^-,0\rangle_{\rho}$, $\Xi_b|J^P=1/2^-,1\rangle_{\rho}$, $\Xi_b|J^P=3/2^-,1\rangle_{\rho}$, $\Xi_b|J^P=3/2^-,2\rangle_{\rho}$, and $\Xi_b|J^P=5/2^-,2\rangle_{\rho}$. As listed in Table~\ref{table2}, the masses of these $\rho$-mode $1P$-wave $\Xi_b$ states fluctuate around $M\sim6.25$ GeV. Firstly, we fix the masses at the predictions in Ref.~\cite{Ortiz-Pacheco:2023kjn}, and study their strong decay properties within the QPC model and ChQM, as collected in Table~\ref{Xiprho}.

\begin{table*}[]
	\caption{\label{Xiprho}The strong decay properties of the $\rho$-mode $1P$-wave $\Xi_b$ states, which masses are taken from the predictions in Ref.~\cite{Ortiz-Pacheco:2023kjn}. $\Gamma_{\text{Total}}$ represents the total decay width, and Expt. denotes the experimental value. The units are MeV.}
	\centering
	\begin{tabular}{ccccccccccccccccc}	
		\hline\hline		&\multicolumn{2}{c}{$\underline{~\Xi_{b}\ket{J^{P}=\frac{1}{2}^{-},0}_{\rho}~}$}&\multicolumn{2}{c}{$\underline{~\Xi_{b}\ket{J^{P}=\frac{1}{2}^{-},1}_{\rho}~}$}
&\multicolumn{2}{c}{$\underline{~\Xi_{b}\ket{J^{P}=\frac{3}{2}^{-},1}_{\rho}~}$}&\multicolumn{2}{c}{$\underline{~\Xi_{b}\ket{J^{P}=\frac{3}{2}^{-},2}_{\rho}~}$}
&\multicolumn{2}{c}{$\underline{~\Xi_{b}\ket{J^{P}=\frac{5}{2}^{-},2}_{\rho}~}$}\\
 &\multicolumn{2}{c}{$M=6243$}&\multicolumn{2}{c}{$M=6258$}&\multicolumn{2}{c}{$M=6249$}&\multicolumn{2}{c}{$M=6265$($\Xi_b(6227)^0  $)}&\multicolumn{2}{c}{$M=6276$($\Xi_b(6227)^0\boldsymbol$)}\\
				\cline{2-11}
Decay width~~&QPC&ChQM&QPC&ChQM&QPC&ChQM&QPC&ChQM&QPC&ChQM\\
		\hline				
		$\Gamma[\Xi_b\pi]$&95.9&0.2&0.0&16.7&0.0&0.0&5.1(3.5)&32.8(22.2)&5.6(3.5)&36.3(22.1)\\
		$\Gamma[\Xi_b'\pi]$&0.0&13.6&70.6&4.3&0.3&1.5&0.7(0.3)&3.6(1.7)&0.3(0.1)&2.0(0.7)\\
        $\Gamma[\Xi_b^{*}\pi]$&0.0&0.0&0.4&2.5&66.6&52.1&0.5(0.2)&2.2(0.9)&0.9(0.3)&5.1(1.7)\\
		$\Gamma[\Lambda_b\bar{K}]$&102.4&0.1&0.0&16.7&0.0&0.0&3.3(1.7)&21.6(11.6)&3.8(1.7)&25.0(11.6)\\
        $\Gamma[\Xi_b\ket{J^{P}=\frac{1}{2}^{-},1}_{\lambda}\pi]$&0.0&0.1&0.5&0.1&0.0&0.0&0.4(0.0)&0.2(0.0)&0.2(0.0)&0.0(0.0)\\
        $\Gamma[\Xi_b\ket{J^{P}=\frac{3}{2}^{-},1}_{\lambda}\pi]$&0.0&0.0&0.0&0.0&0.1&0.0&0.1(0.0)&0.0(0.0)&0.3(0.0)&0.2(0.0)\\
\hline
        $\Gamma_{\text{Total}}$&198.3&14.0&71.5&40.3&67.0&53.6&10.1(5.7)&60.4(36.4)&11.1(5.6)&68.6(36.1)\\
        \hline
	Expt.&\multicolumn{2}{c}{}&\multicolumn{2}{c}{}&\multicolumn{2}{c}{}&\multicolumn{4}{c}{$19^{+5}_{-4}$}\\	
\hline \hline
\end{tabular}	
\end{table*}

In the QPC model, the $\Xi_b|J^P=1/2^-,0\rangle_{\rho}$ may be a broad state with a total decay width of about $\Gamma\simeq198$ MeV. The decay is governed by $\Xi_b\pi$ and $\Lambda_b\bar{K}$ with branching fractions
\begin{equation}
\frac{\Gamma[\Xi_b|J^P=1/2^-,0\rangle_{\rho}\rightarrow \Xi_b \pi]}{\Gamma_{\mathrm{Total}}}\simeq48\%,
\end{equation}
\begin{equation}
\frac{\Gamma[\Xi_b|J^P=1/2^-,0\rangle_{\rho}\rightarrow \Lambda_b \bar{K}]}{\Gamma_{\mathrm{Total}}}\simeq52\%.
\end{equation}

However, in the framework of ChQM, the $\Xi_b|J^P=1/2^-,0\rangle_{\rho}$ might be a moderate state with a width of $\Gamma\simeq14$ MeV, which is about fourteen times smaller than that predicted with the QPC model. Meanwhile, the dominant decay channel become $\Xi_b'\pi$. The predicted branching fraction for this mainly decay channel can reach up to $97\%$.

%Owing to its relatively large width, this state may be difficult to observe experimentally.

The states $\Xi_b|J^P=1/2^-,1\rangle_{\rho}$ and $\Xi_b|J^P=3/2^-,1\rangle_{\rho}$ are probably two moderate states with a comparable width of several tens of MeV. However, their dominant decay channels are different. For the $\Xi_b|J^P=1/2^-,1\rangle_{\rho}$,  under the QPC model the primary decay channel is $\Xi_b'\pi$, whereas under the ChQM the primary decay channels are $\Xi_b\pi$ and $\Lambda_b\bar{K}$. As to $\Xi_b|J^P=3/2^-,1\rangle_{\rho}$, its decay is almost saturated by the $\Xi_b^*\pi$ channel both the QPC model and ChQM. Hence the state may be observed experimentally in the $\Xi_b\pi\pi$ final state via the decay chain: $\Xi_b|J^P=3/2^-,1\rangle_{\rho}\rightarrow \Xi_b^*\pi\rightarrow \Xi_b\pi\pi$.

The other two $1P$-wave $\rho$-mode states $\Xi_b|J^P=3/2^-,2\rangle_{\rho}$ and $\Xi_b|J^P=5/2^-,2\rangle_{\rho}$ are likely not be two very broad states. Although the widths of these two states differ significantly under the QPC model and ChQM, their dominant decay channels are both $\Xi_b\pi$ and $\Lambda_b\bar{K}$. Based on a comparison with the properties of the experimentally observed state $\Xi_b(6227)$, the possibility that the states $\Xi_b|J^P=3/2^-,2\rangle_{\rho}$ and $\Xi_b|J^P=5/2^-,2\rangle_{\rho}$ are candidates for the state $\Xi_b(6227)$ cannot be definitively excluded. Hence, we further take $\Xi_b|J^P=3/2^-,2\rangle_{\rho}$ and $\Xi_b|J^P=5/2^-,2\rangle_{\rho}$ as $\Xi_b(6227)^0$, and list their decay properties in Table~\ref{Xiprho} as well.

It is obtained that the ratio of the two primary decay channels $\Xi_b\pi$ and $\Lambda_b\bar{K}$ is
\begin{equation}
\frac{\Gamma[\Xi_b|J^P=3(5)/2^-,2\rangle_{\lambda}\rightarrow \Lambda_b\bar{K}]}{\Gamma[\Xi_b|J^P=3(5)/2^-,2\rangle_{\lambda}\rightarrow \Xi_b\pi]}\simeq0.5,
\end{equation}
which agrees well with the experimental value measured by the LHCb Collaboration~\cite{LHCb:2018vuc}.
While the total decay width of $\Xi_b|J^P=3/2^-,2\rangle_{\rho}$ and $\Xi_b|J^P=5/2^-,2\rangle_{\rho}$ in the QPC model is about
\begin{equation}
\Gamma_{\mathrm{Total}}\simeq 5.7-5.6~\mathrm{MeV},
\end{equation}
and that in the ChQM is about
\begin{equation}
\Gamma_{\mathrm{Total}}\simeq 36.4-36.1~\mathrm{MeV}.
\end{equation}
The numerical results underestimate or overestimate the experimental value($19^{+5}_{-4}$ MeV) by roughly ten to twenty MeV.
Nonetheless, considering the theoretical uncertainties, the possibility that states $\Xi_b|J^P=3/2^-,2\rangle_{\rho}$ and $\Xi_b|J^P=5/2^-,2\rangle_{\rho}$ are candidates for state $\Xi_b(6227)$ cannot be ruled out. However, the decay properties of $\Xi_b|J^P=3/2^-,2\rangle_{\rho}$ and $\Xi_b|J^P=5/2^-,2\rangle_{\rho}$ are so similar that distinguishing them from each other would be very difficult both in experiment and theory.

\begin{figure}[]
	\centering \epsfxsize=8.5 cm \epsfbox{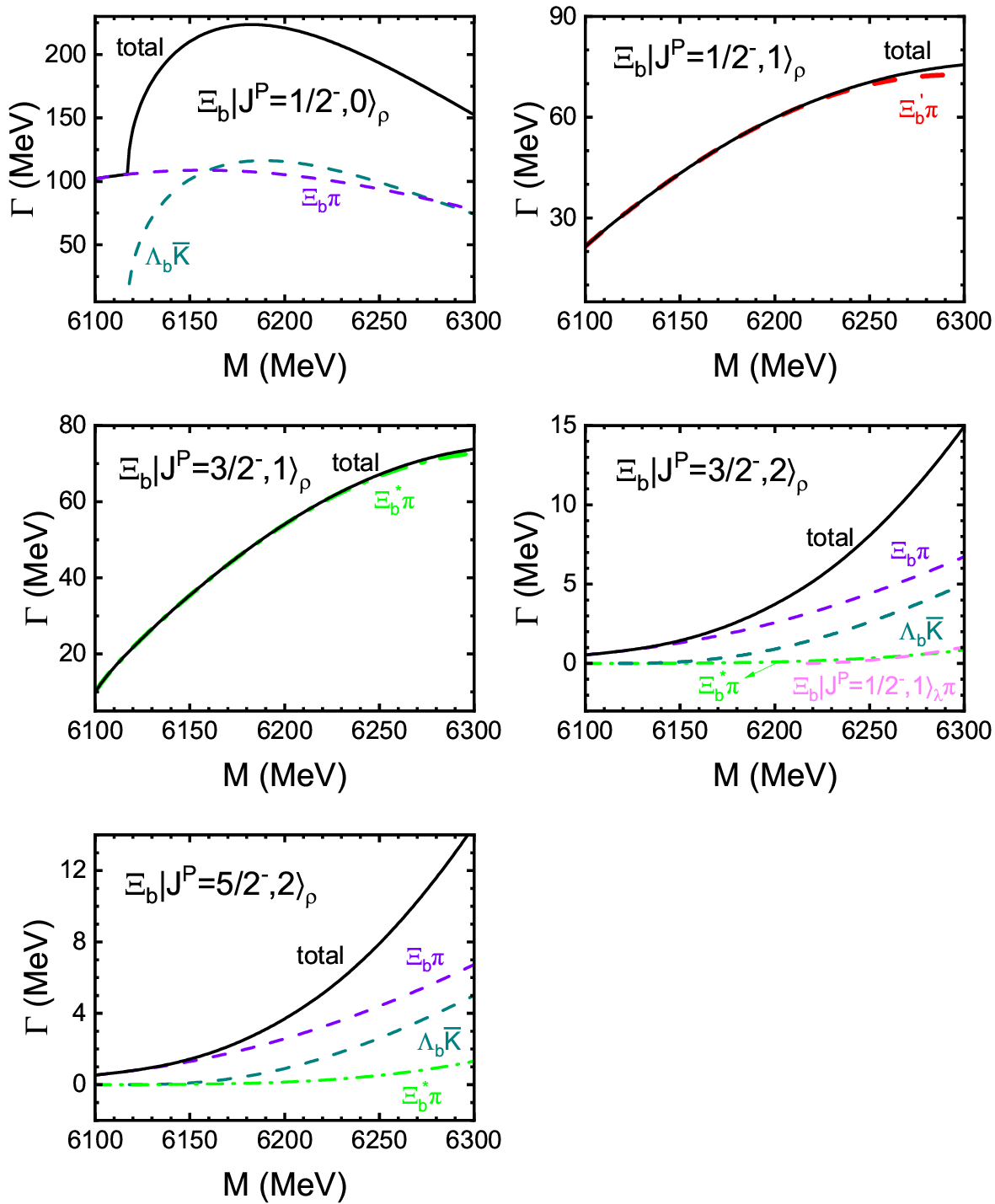}
	\caption{ Partial and total strong decay widths of the $1P$-wave $\rho$-mode $\Xi_b$ excitations as functions of their masses. Some decay channels are too small to show in figure.}
    \label{FIG5}
\end{figure}

Similarly, considering the uncertainty of the masses, we further investigate the strong decay widths of the $1P$-wave $\rho$-mode $\Xi_b$ states with the QPC model as a function of the mass varying in region of $M=(6100-6300)$ MeV in Fig.~\ref{FIG5}. It is shown that the decay properties of the five $1P$-wave $\rho$-mode $\Xi_b$ excitations exhibit a significant dependence on the mass across the defined range.

For the $1P$-wave $\rho$-mode $\Xi_b'$ excitations, there are two states $\Xi_b'|J^P=1/2^-,1\rangle_{\rho}$ and $\Xi_b'|J^P=3/2^-,1\rangle_{\rho}$. For their masses, there are some discussions in theoretical references and we have collected in Table~\ref{table2} as well. From the table, the masses of the $1P$-wave $\rho$-mode $\Xi_b'$ baryons are about $M\sim6.37$ GeV. Fixing their masses on the predictions in Ref.~\cite{Ortiz-Pacheco:2023kjn}, we calculate their strong decay properties, and collect the results in Table~\ref{Xipprho}.

\begin{table}[]
	\caption{\label{Xipprho}The strong decay properties of the $\rho$-mode $1P$-wave $\Xi_b'$ states, which masses are taken from the predictions in Ref.~\cite{Ortiz-Pacheco:2023kjn}. The symbol ``-'' indicates that this decay channel is forbidden. $\Gamma_{\text{Total}}$ represents the total decay width. The units are MeV.}
	\centering
	\begin{tabular}{ccccccccccccccccc}	
		\hline\hline		&\multicolumn{2}{c}{$\underline{~\Xi_{b}'\ket{J^{P}=\frac{1}{2}^{-},1}_{\rho}~}$}&\multicolumn{2}{c}{$\underline{~\Xi_{b}'\ket{J^{P}=\frac{3}{2}^{-},1}_{\rho}~}$}\\
 &\multicolumn{2}{c}{$M=6366$}&\multicolumn{2}{c}{$M=6373$}\\
				\cline{2-5}
Decay width~~&QPC&ChQM&QPC&ChQM\\
		\hline				
		$\Gamma[\Xi_b'\pi]$&33.4&13.6&3.1&11.8\\
		$\Gamma[\Xi_b^{*}\pi]$&4.7&17.8&36.8&24.1\\
        $\Gamma[\Sigma_bK]$&112.4&47.3&1.4&5.5\\
		$\Gamma[\Sigma_b^*K]$&0.8&3.5&109.0&51.5\\
        $\Gamma[\Xi_b'\ket{J^{P}=\frac{1}{2}^{-},0}_{\lambda}\pi]$&0.0&0.0&0.1&0.1\\
       $\Gamma[\Xi_b'\ket{J^{P}=\frac{1}{2}^{-},1}_{\lambda}\pi]$&0.0&0.0&0.0&0.1\\
       $\Gamma[\Xi_b'\ket{J^{P}=\frac{3}{2}^{-},1}_{\lambda}\pi]$&0.0&0.1&0.1&0.2\\
       $\Gamma[\Xi_b'\ket{J^{P}=\frac{3}{2}^{-},2}_{\lambda}\pi]$&0.0&0.0&0.0&0.0\\
       $\Gamma[\Xi_b'\ket{J^{P}=\frac{5}{2}^{-},2}_{\lambda}\pi]$&-&-&0.0&0.0\\
\hline
        $\Gamma_{\text{Total}}$&151.3&82.3&150.5&93.3\\	
\hline \hline
\end{tabular}	
\end{table}

Within the QPC model, the two states $\Xi_b'|J^P=1/2^-,1\rangle_{\rho}$ and $\Xi_b'|J^P=3/2^-,1\rangle_{\rho}$ are broad states with a width of $\Gamma\sim150$ MeV. However, their dominant decay channels are different. For the state $\Xi_b'|J^P=1/2^-,1\rangle_{\rho}$, the mainly decay channel is $\Sigma_bK$, and the corresponding branching fractions can reach up to
\begin{equation}
\frac{\Gamma[\Xi_b'|J^P=1/2^-,1\rangle_{\rho}\rightarrow \Sigma_b K]}{\Gamma_{\mathrm{Total}}}\simeq74\%.
\end{equation}
The large branching ratios indicate the state $\Xi_b'|J^P=1/2^-,1\rangle_{\rho}$ is very likely be observed in the $\Lambda_b\pi K$ final state via the decay chain $\Xi_b'|J^P=1/2^-,1\rangle_{\rho}\rightarrow \Sigma_b K\rightarrow \Lambda_b\pi K$.
Meanwhile, $\Xi_b'|J^P=1/2^-,1\rangle_{\rho}$ may have a sizeable decay rate into $\Xi_b'\pi$, and the predicted branching fraction is
\begin{equation}
\frac{\Gamma[\Xi_b'|J^P=1/2^-,1\rangle_{\rho}\rightarrow \Xi_b'\pi]}{\Gamma_{\mathrm{Total}}}\simeq22\%.
\end{equation}
Hence, the $\Xi_b'|J^P=1/2^-,1\rangle_{\rho}$ may be observed in the $\Xi_b\pi\pi$ final state as well via the decay chain $\Xi_b'|J^P=1/2^-,1\rangle_{\rho}\rightarrow \Xi_b' \pi\rightarrow \Xi_b\pi \pi$.

While, for the other $1P$-wave $\rho$-mode $\Xi_b'$ excitation $\Xi_b'|J^P=3/2^-,1\rangle_{\rho}$, the dominant decay channel is $\Sigma_b^*K$, and the predicted branching fraction is
 \begin{equation}
\frac{\Gamma[\Xi_b'|J^P=3/2^-,1\rangle_{\rho}\rightarrow \Sigma_b^*K]}{\Gamma_{\mathrm{Total}}}\simeq72\%.
\end{equation}
Thus, the state $\Xi_b'|J^P=3/2^-,1\rangle_{\rho}$ may be observed in the $\Lambda_b\pi K$ final state via the decay chain $\Xi_b'|J^P=3/2^-,1\rangle_{\rho}\rightarrow \Sigma_b^* K\rightarrow \Lambda_b\pi K$.
Moreover, the state $\Xi_b'|J^P=3/2^-,1\rangle_{\rho}$ have a sizable decay rate into $\Xi_b^*\pi$ with a branching fraction of
\begin{equation}
\frac{\Gamma[\Xi_b'|J^P=3/2^-,1\rangle_{\rho}\rightarrow \Xi_b^*\pi]}{\Gamma_{\mathrm{Total}}}\simeq24\%.
\end{equation}
Therefore, the decay chain $\Xi_b'|J^P=3/2^-,1\rangle_{\rho}\rightarrow \Xi_b^* \pi\rightarrow \Xi_b\pi \pi$ may be the other optional decay precess to investigate the nature of $\Xi_b'|J^P=3/2^-,1\rangle_{\rho}$.

\begin{figure}[]
	\centering \epsfxsize=8.5 cm \epsfbox{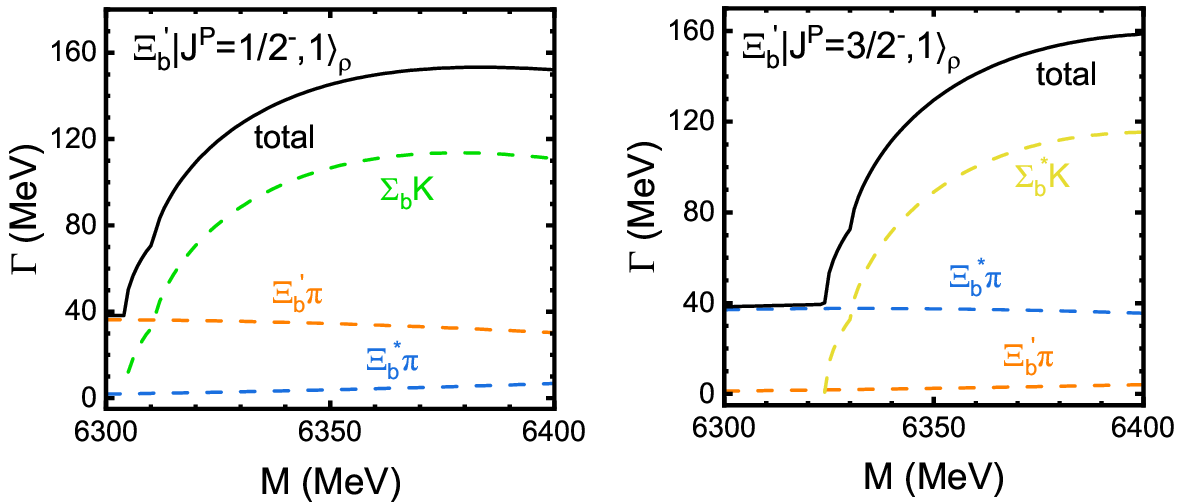}
	\caption{ Partial and total strong decay widths of the $1P$-wave $\rho$-mode $\Xi_b'$ excitations as functions of their masses. Some decay channels are too small to show in figure.}
    \label{FIG6}
\end{figure}

For a comparison, the predicted results with the ChQM are collected in Table~\ref{Xipprho} as well. It is found that the predicted main decay channels with the two models are agreement to each other, however, the predicted total decay widths with the QPC model are about (0.6-0.8) times larger than those with the ChQM.

\begin{table*}[]
	\caption{\label{Xi2SL}The strong decay properties of the $\lambda$-mode $2S$-wave $\Xi_b$ and $\Xi_b'$ states, which masses are taken from the predictions in Ref.~\cite{Ortiz-Pacheco:2023kjn}. The symbol ``-'' indicates that this decay channel is forbidden. $\Gamma_{\text{Total}}$ represents the total decay width. The units are MeV.}
	\centering
	\begin{tabular}{ccccccccccccccccc}	
		\hline\hline		&\multicolumn{2}{c}{$\underline{~\Xi_{b}\ket{J^{P}=\frac{1}{2}^{+},0}_{\lambda}~}$}&\multicolumn{2}{c}{$\underline{~\Xi_{b}'\ket{J^{P}=\frac{1}{2}^{+},1}_{\lambda}~}$}
&\multicolumn{2}{c}{$\underline{~\Xi_{b}'\ket{J^{P}=\frac{3}{2}^{+},1}_{\lambda}~}$}\\
 &\multicolumn{2}{c}{$M=6266$}&\multicolumn{2}{c}{$M=6329$}&\multicolumn{2}{c}{$M=6342$}\\
				\cline{2-7}
Decay width&QPC&ChQM&QPC&ChQM&QPC&ChQM\\
		\hline				
		$\Gamma[\Xi_b\pi]$&-&-&1.5&0.2&1.3&0.4\\
       $\Gamma[\Xi_b'\pi]$&1.5&0.7&2.6&0.7&0.7&0.2\\
		$\Gamma[\Xi_b^{*}\pi]$&2.6&1.4&5.0&0.4&3.3&0.9\\
      $\Gamma[\Lambda_b\bar{K}]$&-&-&2.0&0.0&1.8&0.1\\
        $\Gamma[\Sigma_bK]$&-&-&1.8&1.3&0.8&0.5\\
		$\Gamma[\Sigma_b^*K]$&-&-&-&-&1.4&1.1\\
        $\Gamma[\Xi_b\ket{J^{P}=\frac{1}{2}^{-},1}_{\lambda}\pi]$&-&-&4.8&2.2&0.2&1.5\\
       $\Gamma[\Xi_b\ket{J^{P}=\frac{3}{2}^{-},1}_{\lambda}\pi]$&-&-&0.2&0.4&5.0&0.7\\
\hline
        $\Gamma_{\text{Total}}$&4.1&2.1&17.9&5.2&14.5&5.4\\	
\hline \hline
\end{tabular}	
\end{table*}

We also plot the partial decay widths of the $1P$-wave $\rho$-mode $\Xi_b'$ excitations within the QPC model as a function of the mass in region of $M=(6300-6400)$ MeV. The sensitivities of the decay properties of two states to their masses are shown in Fig.~\ref{FIG6}. Within the masses varied in the region we considered in present work, the total decay widths of the $1P$-wave $\rho$-mode $\Xi_b'$ baryons are about $\Gamma\sim(40-160)$ MeV. We notice that the variation in the total width of $\Xi_b'|J^P=1/2^-,1\rangle_{\rho}$ primarily stems from the mass dependence of its decay channel $\Sigma_bK$, while that of $\Xi_b'|J^P=3/2^-,1\rangle_{\rho}$ primarily stems from the mass dependence of its decay channel $\Sigma_b^*K$.

%\begin{figure}[]
%	\centering \epsfxsize=8 cm \epsfbox{SigmacPLambda.eps}
%	\caption{ Partial and total strong decay widths of the $1P$-wave $\lambda$-mode $\Sigma_c$ excitations as functions of their masses. Some decay channels are too small to show in figure.}
%    \label{FIG2}
%\end{figure}

\begin{figure*}[]
	\centering \epsfxsize=17 cm \epsfbox{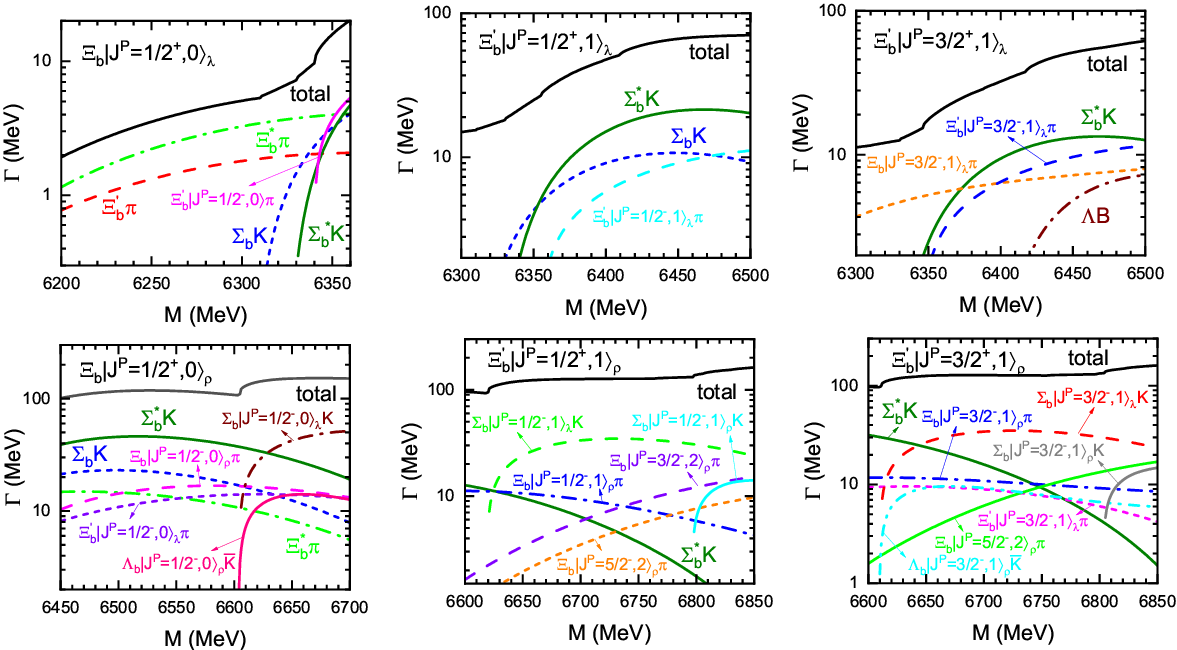}
	\caption{ Partial and total strong decay widths of the $2S$-wave $\Xi_b$ and $\Xi_b'$ excitations as functions of their masses. Some decay channels are too small to show in figure.}
    \label{FIG7}
\end{figure*}

\subsection{$2S$-wave $\lambda$-mode excitations}

For the $2S$-wave $\lambda$-mode $\Xi_b$ baryons, there is only one state $\Xi_b|J^P=1/2^+,0\rangle_{\lambda}$. According to the theoretical predictions by various methods, the mass of the $2S$-wave $\lambda$-mode $\Xi_b$ baryon is about $M\sim6.26$ GeV. Fixing its mass at the prediction in Ref.~\cite{Ortiz-Pacheco:2023kjn}, we calculate its strong decay properties with both the QPC model and ChQM, the theoretical results are listed in Table~\ref{Xi2SL}.

According to our calculations, the $\Xi_b|J^P=1/2^+,0\rangle_{\lambda}$ is a very narrow state with a total decay width of only 2 to 4 MeV. The branching fractions of the mainly decay channels $\Xi_b'\pi$ and $\Xi_b^*\pi$ are
 \begin{equation}
\frac{\Gamma[\Xi_b|J^P=1/2^+,0\rangle_{\lambda}\rightarrow\Xi_b'\pi]}{\Gamma_{\mathrm{Total}}}\simeq(33-37)\%,
\end{equation}
 \begin{equation}
\frac{\Gamma[\Xi_b|J^P=1/2^+,0\rangle_{\lambda}\rightarrow\Xi_b^*\pi]}{\Gamma_{\mathrm{Total}}}\simeq(67-63)\%.
\end{equation}
Due to the narrow decay width, $\Xi_b|J^P=1/2^+,0\rangle_{\lambda}$ has a high probability of being observed experimentally via the decay chain $\Xi_b|J^P=1/2^+,0\rangle_{\lambda}\rightarrow \Xi_b'\pi/\Xi_b^*\pi\rightarrow \Xi_b\pi\pi$.

For the $2S$-wave $\lambda$-mode $\Xi_b'$ baryons, there are two states, which are $\Xi_b'|J^P=1/2^+,1\rangle_{\lambda}$ and $\Xi_b'|J^P=3/2^+,1\rangle_{\lambda}$. We have collected their theoretical masses and possible two-body decay channels in Table~\ref{table2}. From the table, the masses of the $2S$-wave $\lambda$-mode $\Xi_b'$ baryons are about $M\sim6.35$ GeV. Adopting the predicted masses from Ref.~\cite{Ortiz-Pacheco:2023kjn}, we calculate their decay properties and list the theoretical results in Table~\ref{Xi2SL} as well.

 Within the QPC model, the two $2S$-wave $\lambda$-mode $\Xi_b'$ excitations may be narrow states with a total decay widths of approximately a dozen MeV. The $\Xi_b'|J^P=1/2^+,1\rangle_{\lambda}$  predominantly decays to $\Xi_b^*\pi$ and $\Xi_b|J^P=1/2^-,1\rangle_{\lambda}\pi$. The branching fractions for the two mainly decay channels are
 \begin{equation}
\frac{\Gamma[\Xi_b'|J^P=1/2^+,1\rangle_{\lambda}\rightarrow\Xi_b^*\pi]}{\Gamma_{\mathrm{Total}}}\simeq30\%,
\end{equation}
 \begin{equation}
\frac{\Gamma[\Xi_b'|J^P=1/2^+,1\rangle_{\lambda}\rightarrow \Xi_b|J^P=1/2^-,1\rangle_{\lambda}\pi]}{\Gamma_{\mathrm{Total}}}\simeq27\%.
\end{equation}
While, the decay of $\Xi_b'|J^P=3/2^+,1\rangle_{\lambda}$ is governed by $\Xi_b|J^P=3/2^-,1\rangle_{\lambda}\pi$ with branching fraction
\begin{equation}
\frac{\Gamma[\Xi_b'|J^P=3/2^+,1\rangle_{\lambda}\rightarrow \Xi_b|J^P=3/2^-,1\rangle_{\lambda}\pi]}{\Gamma_{\mathrm{Total}}}\simeq34\%.
\end{equation}
Meanwhile, the decay channel $\Xi_b^*\pi$ also has a significant branching fraction, which is
\begin{equation}
\frac{\Gamma[\Xi_b'|J^P=3/2^+,1\rangle_{\lambda}\rightarrow\Xi_b^*\pi]}{\Gamma_{\mathrm{Total}}}\simeq23\%.
\end{equation}

The decay properties predicted by the ChQM for the $2S$-wave $\lambda$-mode $\Xi_b'$ states are somewhat different from those predicted by the QPC model. Within the ChQM, the total decay width of the $2S$-wave $\lambda$-mode $\Xi_b'$ states is about 0.4 times of that calculated by the QPC model. Meanwhile, the dominant decay channels of $\Xi_b'|J^P=1/2^+,1\rangle_{\lambda}$ are $\Sigma_bK$ and $\Xi_b|J^P=1/2^-,1\rangle_{\lambda}\pi$. And the mainly decay modes of  $\Xi_b'|J^P=3/2^+,1\rangle_{\lambda}$ are $\Sigma^*_bK$ and $\Xi_b|J^P=1/2^-,1\rangle_{\lambda}\pi$.

Considering the uncertainty of the masses of the $2S$-wave $\lambda$-mode states, we further investigate the strong decay widths of the $2S$-wave $\lambda$-mode states with the QPC model as a function of the mass in Fig.~\ref{FIG7}. The mass of the $2S$-wave $\lambda$-mode $\Xi_b$ state varies in the region of $M=(6200-6360)$ MeV, and those of the $2S$-wave $\lambda$-mode $\Xi_b'$ states vary in the region of $M=(6300-6500)$ MeV. From the figure, it is obtained that with the masses increasing the $\Sigma_b^*K$ becomes the dominant decay channel for the $2S$-wave $\lambda$-mode $\Xi_b'$ resonances. Meanwhile, the final state containing the $1P$-wave charmed baryon will become more and more important as well.

\subsection{$2S$-wave $\rho$-mode excitations}

\begin{table*}[]
	\caption{\label{Xi2Sr}The strong decay properties of the $\rho$-mode $2S$-wave $\Xi_b$ and $\Xi_b'$ states, which masses are 150 MeV smaller than the predictions in Ref.~\cite{Garcia-Tecocoatzi:2023btk}. The symbol ``-'' indicates that this decay channel is forbidden. $\Gamma_{\text{Total}}$ represents the total decay width. The units are MeV.}
	\centering
	\begin{tabular}{ccccccccccccccccc}	
		\hline\hline		&\multicolumn{2}{c}{$\underline{~\Xi_{b}\ket{J^{P}=\frac{1}{2}^{+},0}_{\rho}~}$}&\multicolumn{2}{c}{$\underline{~\Xi_{b}'\ket{J^{P}=\frac{1}{2}^{+},1}_{\rho}~}$}
&\multicolumn{2}{c}{$\underline{~\Xi_{b}'\ket{J^{P}=\frac{3}{2}^{+},1}_{\rho}~}$}\\
 &\multicolumn{2}{c}{$M=6549$}&\multicolumn{2}{c}{$M=6668$}&\multicolumn{2}{c}{$M=6697$}\\
				\cline{2-7}
Decay width&QPC&ChQM&QPC&ChQM&QPC&ChQM\\
		\hline				
		$\Gamma[\Xi_b\pi]$&-&-&0.5&6.6&0.2&8.3\\
       $\Gamma[\Xi_b'\pi]$&6.2&0.9&4.1&0.4&0.8&0.4\\
       $\Gamma[\Xi_b^*\pi]$&13.2&2.1&2.4&0.3&4.7&1.0\\

       $\Gamma[\Xi_b\eta]$&-&-&1.7&0.5&1.3&0.7\\
      $\Gamma[\Xi_b'\eta]$&-&-&4.0&1.0&1.0&0.1\\
       $\Gamma[\Xi_b^*\eta]$&-&-&2.0&3.1&5.0&1.0\\

      $\Gamma[\Lambda_b\bar{K}]$&-&-&0.1&7.8&0.0&8.6\\
      $\Gamma[\Sigma_bK]$&21.3&0.5&14.3&2.9&2.7&1.5\\
       $\Gamma[\Sigma_b^*K]$&44.7&1.4&8.4&8.2&16.4&4.9\\

      $\Gamma[\Lambda B]$&-&-&0.0&-&0.0&-\\
      $\Gamma[\Sigma B]$&-&-&0.0&-&0.0&-\\
      $\Gamma[\Sigma^* B]$&-&-&0.0&-&0.0&-\\

      $\Gamma[\Xi B_s]$&-&-&-&-&0.0&-\\
      $\Gamma[\Xi_b\ket{J^{P}=\frac{1}{2}^{-},1}_{\lambda}\pi/\eta]$&-/-&-/-&3.1/1.6&59.5/12.6&0.5/0.0&0.1/0.0\\
      $\Gamma[\Xi_b\ket{J^{P}=\frac{3}{2}^{-},1}_{\lambda}\pi/\eta]$&-/-&-/-&1.1/0.0&0.4/0.0&3.2/2.0&9.7/21.4\\
    $\Gamma[\Xi_b\ket{J^{P}=\frac{1}{2}^{-},0}_{\rho}\pi]$&16.4&24.1&0.0&0.0&0.0&0.0\\
    $\Gamma[\Xi_b\ket{J^{P}=\frac{1}{2}^{-},1}_{\rho}\pi]$&0.0&0.0&10.2&12.3&1.7&6.9\\
    $\Gamma[\Xi_b\ket{J^{P}=\frac{3}{2}^{-},1}_{\rho}\pi]$&0.0&0.0&2.8&8.2&10.9&17.8\\
    $\Gamma[\Xi_b\ket{J^{P}=\frac{3}{2}^{-},2}_{\rho}\pi]$&0.8&3.9&4.2&17.7&2.8&11.8\\
    $\Gamma[\Xi_b\ket{J^{P}=\frac{5}{2}^{-},2}_{\rho}\pi]$&0.9&3.2&2.4&8.6&5.9&11.1\\

$\Gamma[\Xi_b'\ket{J^{P}=\frac{1}{2}^{-},0}_{\lambda}\pi]$&12.9&1.6&0.0&0.0&0.0&0.0\\
$\Gamma[\Xi_b'\ket{J^{P}=\frac{1}{2}^{-},1}_{\lambda}\pi]$&0.0&0.4&9.3&0.7&0.2&0.2\\
$\Gamma[\Xi_b'\ket{J^{P}=\frac{3}{2}^{-},1}_{\lambda}\pi]$&0.0&4.1&0.4&0.0&9.0&0.8\\
$\Gamma[\Xi_b'\ket{J^{P}=\frac{3}{2}^{-},2}_{\lambda}\pi]$&0.2&2.6&0.7&1.0&0.4&0.4\\
$\Gamma[\Xi_b'\ket{J^{P}=\frac{5}{2}^{-},2}_{\lambda}\pi]$&0.3&0.5&0.4&0.5&0.9&0.9\\
  $\Gamma[\Xi_b'\ket{J^{P}=\frac{1}{2}^{-},1}_{\rho}\pi]$&-&-&5.4&8.2&0.9&2.9\\
  $\Gamma[\Xi_b'\ket{J^{P}=\frac{3}{2}^{-},1}_{\rho}\pi]$&-&-&0.9&10.4&6.3&10.9\\

$\Gamma[\Lambda_b\ket{J^{P}=\frac{1}{2}^{-},1}_{\lambda}\bar{K}]$&-&-&3.3&20.6&0.6&0.0\\
$\Gamma[\Lambda_b\ket{J^{P}=\frac{3}{2}^{-},1}_{\lambda}\bar{K}]$&-&-&1.2&0.1&3.4&7.7\\
    $\Gamma[\Lambda_b\ket{J^{P}=\frac{1}{2}^{-},0}_{\rho}\bar{K}]$&-&-&0.0&0.0&0.0&0.0\\
    $\Gamma[\Lambda_b\ket{J^{P}=\frac{1}{2}^{-},1}_{\rho}\bar{K}]$&-&-&9.3&14.8&0.4&2.1\\
    $\Gamma[\Lambda_b\ket{J^{P}=\frac{3}{2}^{-},1}_{\rho}\bar{K}]$&-&-&0.4&0.8&9.3&14.4\\
    $\Gamma[\Lambda_b\ket{J^{P}=\frac{3}{2}^{-},2}_{\rho}\bar{K}]$&-&-&0.4&2.8&0.6&3.8\\
    $\Gamma[\Lambda_b\ket{J^{P}=\frac{5}{2}^{-},2}_{\rho}\bar{K}]$&-&-&0.1&0.5&1.0&2.7\\

 $\Gamma[\Sigma_b\ket{J^{P}=\frac{1}{2}^{-},0}_{\lambda}K]$&-&-&0.0&0.0&0.0&0.0\\
 $\Gamma[\Sigma_b\ket{J^{P}=\frac{1}{2}^{-},1}_{\lambda}K]$&-&-&28.8&1.0&0.2&0.8\\
 $\Gamma[\Sigma_b\ket{J^{P}=\frac{3}{2}^{-},1}_{\lambda}K]$&-&-&0.2&0.0&34.1&1.3\\
 $\Gamma[\Sigma_b\ket{J^{P}=\frac{3}{2}^{-},2}_{\lambda}K]$&-&-&0.2&0.0&0.3&0.4\\
 $\Gamma[\Sigma_b\ket{J^{P}=\frac{5}{2}^{-},2}_{\lambda}K]$&-&-&0.1&0.1&0.5&0.4\\

$\Gamma[\Xi_b\ket{J^{P}=\frac{3}{2}^{+},2}_{\lambda\lambda/\rho\rho}\pi]$&-/-&-/-&0.2/0.1&0.8/1.9&0.0/0.0&0.3/1.0\\
$\Gamma[\Xi_b\ket{J^{P}=\frac{5}{2}^{+},2}_{\lambda\lambda/\rho\rho}\pi]$&-/-&-/-&0.0/0.0&26.4/0.0&0.2/0.1&1.3/4.6\\

$\Gamma[\Xi_b'\ket{J^{P}=\frac{1}{2}^{+},1}_{\lambda\lambda}\pi]$&-&-&0.0&0.0&0.0&0.0\\
$\Gamma[\Xi_b'\ket{J^{P}=\frac{3}{2}^{+},1}_{\lambda\lambda}\pi]$&-&-&0.0&0.1&0.0&0.0\\
$\Gamma[\Xi_b'\ket{J^{P}=\frac{3}{2}^{+},2}_{\lambda\lambda}\pi]$&-&-&0.1&0.0&0.0&0.0\\
$\Gamma[\Xi_b'\ket{J^{P}=\frac{5}{2}^{+},2}_{\lambda\lambda}\pi]$&-&-&0.0&1.4&0.2&0.0\\
$\Gamma[\Xi_b'\ket{J^{P}=\frac{5}{2}^{+},3}_{\lambda\lambda}\pi]$&-&-&0.0&4.5&0.0&0.0\\
$\Gamma[\Xi_b'\ket{J^{P}=\frac{7}{2}^{+},3}_{\lambda\lambda}\pi]$&-&-&0.0&0.0&0.0&5.0\\

$\Gamma[\Lambda_b\ket{J^{P}=\frac{3}{2}^{+},2}_{\lambda\lambda}\bar{K}]$&-&-&-&-&0.0&0.0\\
$\Gamma[\Lambda_b\ket{J^{P}=\frac{5}{2}^{+},2}_{\lambda\lambda}\bar{K}]$&-&-&-&-&0.2&0.0\\

\hline
        $\Gamma_{\text{Total}}$&116.9&45.3&124.4&246.7&127.9&167.2\\	
\hline \hline
\end{tabular}	
\end{table*}

For the $2S$-wave $\rho$-mode $\Xi_b$ baryons, there is only one state $\Xi_b|J^P=1/2^+,0\rangle_{\rho}$ as well according to the symmetry of the wave function. For its mass, there are a few discussions in theoretical references and we have collected in Table~\ref{table2}. From the table, the mass of the $2S$-wave $\rho$-mode $\Xi_b$ baryon in Ref.~\cite{Garcia-Tecocoatzi:2023btk} is about 6.70 GeV. While, a comparison between the Literature~\cite{Garcia-Tecocoatzi:2023btk} and other studies regarding the calculated mass values for the $2S$-wave $\lambda$-mode states reveals that Literature~\cite{Garcia-Tecocoatzi:2023btk} overestimates the mass of the $2S$-wave $\rho$-mode $\Xi_b$ baryon by approximately 150 MeV. Hence, we fix the mass of $\Xi_b|J^P=1/2^+,0\rangle_{\rho}$ at $M=6549$ MeV, and collect its partial decay widths in Table~\ref{Xi2Sr}.

From the table, $\Xi_b|J^P=1/2^+,0\rangle_{\rho}$ has a total width of about $\Gamma\sim117$ MeV, and mainly decays into $\Sigma_b^*K$. The corresponding branching fraction is
\begin{equation}
\frac{\Gamma[\Xi_b|J^P=1/2^+,0\rangle_{\rho}\rightarrow\Sigma_b^*K]}{\Gamma_{\mathrm{Total}}}\simeq38\%.
\end{equation}
Meanwhile, the branching ratios to $\Sigma_bK$, $\Xi_b^*\pi$, $\Xi_b|J^P=1/2^-,0\rangle_{\rho}\pi$ and  $\Xi_b'|J^P=1/2^-,0\rangle_{\lambda}\pi$ are also substantial, with values of 18\%,11\%,14\%, and 11\%,
respectively.

However, according to the calculations within the ChQM, the total decay width of $\Xi_b|J^P=1/2^+,0\rangle_{\rho}$ is $\Gamma\simeq45.3$ MeV, which is about 0.4 times of that predicted by the QPC model. Meanwhile, the dominant decay channel is $\Xi_b|J^P=1/2^-,0\rangle_{\rho}\pi$, and the branching fraction for this dominant decay channel is 53\%.

As to the $2S$-wave $\rho$-mode $\Xi_b'$ baryons, there are two states: $\Xi_b'|J^P=1/2^+,1\rangle_{\rho}$ and $\Xi_b'|J^P=3/2^+,1\rangle_{\rho}$. Their theoretical masses and possible two-body decay channels are listed in Table~\ref{table2}. We fix their masses at $M=6668$ MeV and 6697 MeV, respectively. These values are 150 MeV smaller than the predictions in Ref.~\cite{Garcia-Tecocoatzi:2023btk}, and we collect their decay properties in Table~\ref{Xi2Sr} as well.

From the table, within the QPC model $\Xi_b'|J^P=1/2^+,1\rangle_{\rho}$ and $\Xi_b'|J^P=3/2^+,1\rangle_{\rho}$ have a comparable decay width of about $\Gamma\sim120$ MeV. However, their dominant decay channels are different. $\Xi_b'|J^P=1/2^+,1\rangle_{\rho}$ decays predominantly to the $\Sigma_b|J^P=1/2^-,1\rangle_{\lambda}K$ channel, and the predicted branching fraction is
\begin{equation}
\frac{\Gamma[\Xi_b'|J^P=1/2^+,1\rangle_{\rho}\rightarrow\Sigma_b|J^P=1/2^-,1\rangle_{\lambda}K]}{\Gamma_{\mathrm{Total}}}\simeq23\%.
\end{equation}
Furthermore, the $\Xi_b'|J^P=1/2^+,1\rangle_{\rho}$ state also decays to the $\Sigma_bK$ and $\Xi_b|J^P=1/2^-,1\rangle_{\rho}\pi$ channels with considerable branching ratios, which are about 11\% and 8\%, respectively.

For the $\Xi_b'|J^P=3/2^+,1\rangle_{\rho}$ state, its dominant decay channel is $\Sigma_b|J^P=3/2^-,1\rangle_{\lambda}K$. The branching fraction is predicted to be
\begin{equation}
\frac{\Gamma[\Xi_b'|J^P=3/2^+,1\rangle_{\rho}\rightarrow\Sigma_b|J^P=3/2^-,1\rangle_{\lambda}K]}{\Gamma_{\mathrm{Total}}}\simeq27\%.
\end{equation}
Additionally, the branching ratios for the decays of the $\Xi_b'|J^P=3/2^+,1\rangle_{\rho}$ state to the $\Sigma_b^*K$ and $\Xi_b|J^P=3/2^-,1\rangle_{\rho}\pi$ channels are also sizable, at 13\% and 9\%, respectively.

However, within the ChQM, the total decay width of $\Xi_b'|J^P=1/2^+,1\rangle_{\rho}$ is about 120 MeV larger than the that of prediction via the QPC model. Furthermore, the dominant decay channel of $\Xi_b'|J^P=1/2^+,1\rangle_{\rho}$ is $\Xi_b|J^P=1/2^-,1\rangle_{\lambda}\pi$, with branching fraction being about 24\%. As to the $\Xi_b'|J^P=3/2^+,1\rangle_{\rho}$ state, the total decay width calculated by the ChQM is comparable to that predicted by the QPC model. While, the main decay channels predicted by the two models are different. In the ChQM, the mainly channels for $\Xi_b'|J^P=3/2^+,1\rangle_{\rho}$ are $\Xi_b|J^P=3/2^-,1\rangle_{\lambda}\eta$ and $\Xi_b|J^P=3/2^-,1\rangle_{\rho}\pi$, with their respective branching ratios being 13\% and 11\%.

In addition, considering the uncertainty of the masses of the $2S$-wave states, we further investigate the strong decay widths of $2S$-wave $\Xi_b$ and $\Xi_b'$ states with the QPC model a function of the mass in Fig.~\ref{FIG7}. The mass of the $2S$-wave $\rho$-mode $\Xi_b$ state varies in the region of $M=(6450-6700)$ MeV, and those of the $2S$-wave $\rho$-mode $\Xi_b'$ states vary in the region of $M=(6600-6850)$ MeV. From the figure, we obtain that for the $2S$-wave $\rho$-mode states, apart from ground states such as $\Sigma_bK$ and $\Sigma_b^*K$, the decay channels involving $P$-wave final states also play a significant role.

It should be mentioned that the QPC model and ChQM have distinct physical interpretations. Our theoretical results also exhibit a degree of model dependence. Nevertheless, both models have achieved significant success in describing the strong decays of hadron system. Therefore, the theoretical results presented here await experimental verification. Furthermore, it should be clarified that the form factor $e^{-\frac{\mathbf{p}^2}{2\Lambda^2}}$ is introduced to suppress the nonphysical contributions in the high momentum region. Although the effect of the form factor on the partial decay width becomes more pronounced with increasing momentum, its influence remains limited and does not affect the main physical results. Meanwhile, we also note the dependence of the decay width on the cut-off parameter. Varying the cutoff parameter by approximately 15$\%$ causes a very small change in the width, with only a few of the broader states exhibit a variation on the order of 3 to 4 MeV.

\section{Summary}

In this work, we systematically investigated the two-body strong decays of low-lying $1P$- and $2S$-wave
$\Xi_b$ and $\Xi_b'$ excitations in the QPC model within the $j$-$j$ coupling scheme. For a comparison, we also
give our predictions within the chiral quark model. The main theoretical results are summarized as follows.

For the $1P$-wave $\lambda$-mode $\Xi_b$ states, $\Xi_b|J^{P}=1/2^-,1\rangle_{\lambda}$ is a very narrow state with a total decay width of about $\Gamma\simeq(1-2)$ MeV, and dominantly decays into the $\Xi_b'\pi$ channel. The other $1P$-wave $\lambda$-mode $\Xi_b$ state $\Xi_b|J^{P}=3/2^-,1\rangle_{\lambda}$ is likely a narrower state with a total decay width approximately half that of  $\Xi_b|J^{P}=1/2^-,1\rangle_{\lambda}$. Its decay width is almost saturated by $\Xi_b^*\pi$. Combining the measured masses and decay properties of the observed states $\Xi_b(6087)$ and $\Xi_b(6095/6100)$,  $\Xi_b|J^{P}=1/2^-,1\rangle_{\lambda}$ and $\Xi_b|J^{P}=3/2^-,1\rangle_{\lambda}$ are good candidates for the states $\Xi_b(6087)$ and $\Xi_b(6095/6100)$, respectively.

For the $1P$-wave $\lambda$-mode $\Xi_b'$ states, within the QPC model $\Xi_b'|J^{P}=1/2^-,0\rangle_{\lambda}$ is likely a moderate state with a decay width of $\Gamma\sim95$ MeV. Meanwhile, the dominant channels are $\Lambda_b\bar{K}$ and $\Xi_b\pi$. The total widths of $\Xi_b'|J^{P}=1/2^-,1\rangle_{\lambda}$ and $\Xi_b'|J^{P}=3/2^-,1\rangle_{\lambda}$ are comparable in the QPC model, both in the range of $\Gamma\sim(20-30)$ MeV. However, their dominant decay channels are different. $\Xi_b'|J^{P}=1/2^-,1\rangle_{\lambda}$ mainly decays into $\Xi_b'\pi$, while $\Xi_b'|J^{P}=3/2^-,1\rangle_{\lambda}$ mainly decays into $\Xi_b^*\pi$. The dominant decay channels predicted by the ChQM for the three $1P$-wave $\lambda$-mode states mentioned above are highly consistent with the predictions of the QPC model. However, the total decay width for the three states in the ChQM is about half of that in the QPC model. Both $\Xi_b'|J^{P}=3/2^-,2\rangle_{\lambda}$ and $\Xi_b'|J^{P}=5/2^-,2\rangle_{\lambda}$ have similar decay properties. They have large decay rates into the $\Xi_b\pi$ and $\Lambda_b\bar{K}$ channels. Their total decay widths within the ChQM are about 24 MeV, which are good agreement with the measured the total width of the observed state $\Xi_b(6227)$. Hence, $\Xi_b'|J^{P}=3/2^-,2\rangle_{\lambda}$ and $\Xi_b'|J^{P}=5/2^-,2\rangle_{\lambda}$ could be good candidates for the state $\Xi_b(6227)$.

For the $1P$-wave $\rho$-mode $\Xi_b$ states, the decay properties of the two states $\Xi_b|J^{P}=1/2^-,0\rangle_{\rho}$ and $\Xi_b|J^{P}=1/2^-,1\rangle_{\rho}$ have a strong model dependency. In the QPC model, $\Xi_b|J^{P}=1/2^-,0\rangle_{\rho}$ has a broad width of $\Gamma\sim198$ MeV, and mainly decays into the $\Xi_b\pi$ and $\Lambda_b\bar{K}$ channels. $\Xi_b|J^{P}=1/2^-,1\rangle_{\rho}$ may be a moderate state with a total decay width of $\Gamma\sim72$ MeV. Its decay width is almost saturated by the $\Xi_b'\pi$ channel. However, in the ChQM, $\Xi_b|J^{P}=1/2^-,0\rangle_{\rho}$ is predicted to be a narrow state, and its decay is governed by the $\Xi_b'\pi$ channel. $\Xi_b|J^{P}=1/2^-,1\rangle_{\rho}$ still have a moderate width, while the dominant decay channels are now the $\Xi_b\pi$ and $\Lambda_b\bar{K}$ channels. The state $\Xi_b|J^{P}=3/2^-,1\rangle_{\rho}$ dominantly decay into $\Xi_b^*\pi$, and probably has a moderate width of $\Gamma\sim60$ MeV.
$\Xi_b|J^{P}=3/2^-,2\rangle_{\rho}$ and $\Xi_b|J^{P}=5/2^-,2\rangle_{\rho}$ have similar decay properties, and mainly decay into $\Xi_b\pi$ and $\Lambda_b\bar{K}$. Their total decay widths in the QPC model and ChQM are about 6 MeV and 36 MeV, respectively, which underestimate or overestimate the experimental width of the state $\Xi_b(6227)$. While, considering the theoretical uncertainties, the possibility that states $\Xi_b|J^{P}=3/2^-,2\rangle_{\rho}$ and $\Xi_b|J^{P}=5/2^-,2\rangle_{\rho}$ are candidates for $\Xi_b(6227)$ cannot be ruled out.

For the $1P$-wave $\rho$-mode $\Xi_b'$ states, within the QPC model $\Xi_b'|J^{P}=1/2^-,1\rangle_{\rho}$ and $\Xi_b'|J^{P}=3/2^-,1\rangle_{\rho}$ are broad states with a comparable width of $\Gamma\sim151$ MeV. However, the dominant decay channel of $\Xi_b'|J^{P}=1/2^-,1\rangle_{\rho}$ is $\Sigma_bK$, and that of $\Xi_b'|J^{P}=3/2^-,1\rangle_{\rho}$ is $\Sigma_b^*K$. With the ChQM, the predicted main decay channels for the two $1P$-wave $\rho$-mode $\Xi_b'$ states are agreement to the predictions in the QPC model. While, the total decay widths are about $\Gamma\sim(82-93)$ MeV, which are about (0.6-0.8) times smaller than those of calculated by the QPC model.

For the $2S$-wave $\lambda$-mode $\Xi_b$ state $\Xi_b|J^P=1/2^+,0\rangle_{\lambda}$, similar decay properties are predicted within the two models. $\Xi_b|J^P=1/2^+,0\rangle_{\lambda}$ may be a very narrow state with a total decay width of only 2 to 4 MeV, and has a high probability of being observed experimentally via the decay chain $\Xi_b|J^P=1/2^+,0\rangle_{\lambda}\rightarrow \Xi_b'\pi/\Xi_b^*\pi\rightarrow \Xi_b\pi\pi$.

For the $2S$-wave $\lambda$-mode $\Xi_b'$ states, their total decay widths are approximately a dozen MeV within the QPC model. $\Xi_b'|J^P=1/2^+,1\rangle_{\lambda}$ predominantly decays to $\Xi_b^*\pi$ and $\Xi_b|J^P=1/2^-,1\rangle_{\lambda}\pi$. While, $\Xi_b'|J^P=3/2^+,1\rangle_{\lambda}$ predominantly decays to $\Xi_b^*\pi$ and $\Xi_b|J^P=3/2^-,1\rangle_{\lambda}\pi$. However, within the ChQM, the total decay widths of the two $2S$-wave $\lambda$-mode $\Xi_b'$ states are about 0.4 times of those predicted by the QPC model.  Meanwhile, the dominant decay channels of $\Xi_b'|J^P=1/2^+,1\rangle_{\lambda}$ are $\Xi_b'\pi$ and $\Xi_b|J^P=1/2^-,1\rangle_{\lambda}\pi$. And the mainly decay modes of  $\Xi_b'|J^P=3/2^+,1\rangle_{\lambda}$ are $\Xi_b^*\pi$ and $\Xi_b|J^P=1/2^-,1\rangle_{\lambda}\pi$.

As to the $2S$-wave $\rho$-mode states, $\Xi_b|J^P=1/2^+,0\rangle_{\rho}$ has a total decay width of dozens to one hundred MeV.
$\Xi_b'|J^P=1/2^+,1\rangle_{\rho}$ and $\Xi_b'|J^P=3/2^+,1\rangle_{\rho}$ may be two broad states, with a total decay width of over one hundred to two hundred MeV. Meanwhile, according to our calculations, in addition to decay channels $\Sigma_b^{(*)}K$ and $\Xi_b^*\pi$, two-body decay channels involving $1P$-wave final states also play a significant role in experimental detection.

%It should be mentioned that the QPC model and ChQM are based on different physical pictures. Within the QPC model, the strong decay takes place via the creation of quark-antiquark pair from the vacuum with quantum number $0^{++}$. Meanwhile, the final meson is taken as a particle with an internal structure. For ChQM, the strong decay takes place via the creation of quark-antiquark pair from the vacuum as well,  while the final meson is taken as a point particle which coupling to quarks satisfies the chiral symmetry and consistent with the spirit of QCD. The two models both achieve great success in describing the strong decays of hadron system. Hence, the results obtained in this work await experimental verification in future.

\section*{Acknowledgements }

This work is supported by the National Natural Science Foundation of China under Grants No.12005013.

\bibliography{Xib}

%\begin{thebibliography}{99}

%%\cite{Ni:2023lvx}
%\bibitem{Ni:2023lvx}
%R.~H.~Ni, J.~J.~Wu and X.~H.~Zhong,
%Unified unquenched quark model for heavy-light mesons with chiral dynamics,
%Phys. Rev. D \textbf{109} no.11, 116006(2024).
%%doi:10.1103/PhysRevD.109.116006
%%[arXiv:2312.04765 [hep-ph]].
%%19 citations counted in INSPIRE as of 15 Dec 2025

%\end{thebibliography}
	\end{document}